\documentclass[11pt,preprint,flushrt]{aastex}
\shorttitle{Advanced Exposure-Time Calculations}
\slugcomment{Revision: 1.9, Date: 2001/09/12 15:07:13, Accepted to
PASP}
\begin{document}

\def\shah{{\rm III}}
\def\eqq#1{Equation~(\ref{#1})}
\def\Var#1{{\rm Var}(#1)}
\def\etal{{\it et al.}}
\def\micron{\hbox{$\mu$m}}

\title{Advanced Exposure-Time Calculations:  Undersampling, Dithering,
Cosmic Rays, Astrometry, and Ellipticities}
\author{Gary Bernstein}
\affil{Dept. of Astronomy, University of Michigan,
830 Dennison Bldg, Ann Arbor, MI 48109, 
garyb@astro.lsa.umich.edu}

\begin{abstract}
The familiar tools of Fourier analysis and Fisher matrices are applied
to derive the uncertainties on photometric, astrometric, and
weak-lensing measurements of stars and galaxies in real astronomical
images.  Many effects or functions that are ignored in basic
exposure-time calculators can be included in this framework:
pixels of size comparable to the stellar image; undersampled and
dithered exposures; cosmic-ray hits; intrapixel sensitivity
variations; positional and ellipticity errors as well as photometric
errors. I present a formalism and a {\tt C++} implementation of these
methods.  As examples of their use, I answer some commonly arising
questions about 
imaging strategies:  What amount of dithering is ideal?  What pixel
size optimizes the productivity of a camera?  Which is more
efficient---space-based or ground-based observing?  
\end{abstract}


\section{Introduction}
A basic exercise in the design of any astronomical camera or observing
program is the estimation of the expected uncertainties, typically in
the form of the photometric $S/N$ ratio for a source of a given
angular size in a given exposure time.
Calculation of the $S/N$ from aperture photometry is straightforward
once the characteristics of the source, sky background, telescope, and
detector have been ascertained.  Derivations of these calculations,
and web-based forms to perform them, can be found, for example, 
in the online documentation for the {\em HST} and for NOAO telescopes.  

The common aperture-photometry formulae give a good rough estimate of
the expected performance, but do not address
several issues that are critical to optimizing a telescope design or
observing-program design.  The goal of this paper is to demonstrate a means
to incorporate the following issues into an exposure-time analysis:
\begin{itemize}
\item Aperture photometry is not optimal; for unresolved sources,
point-spread-function (PSF) fitting techniques are optimal.  What is
the accuracy of PSF-fitting photometry, especially for
diffraction-limited point-spread functions (PSFs) from obscured
circular apertures?
\item The aperture formulae assume pixels either much smaller than or
much larger than the optical PSF.  The intermediate case is more
common, and must be handled by creating an ``effective PSF'' which
includes the pixelization.
\item What are the errors for positional measurements on point sources
(astrometry), galaxy magnitudes/colors (photometric redshifts), and
galaxy ellipticities (weak lensing) under optimal analyses?
\item How does undersampling or sub-pixel dithering affect accuracy?
\item How do intrapixel sensitivity variations---{\it e.g.} the
``picture frame'' effect typical of HgCdTe detector pixels---affect
these measurements?  Likewise, what about charge diffusion within the
detector? 
\item How does one quantitatively assess the impact of cosmic-ray hits
without having to produce Monte-Carlo images?
\item How does uncertainty in the source position affect photometric
estimates? 
\end{itemize}

The impetus for this work is to predict precisely the performance of
various configurations of the proposed {\em Supernova Acceleration
Probe}\footnote{{\tt http://snap.lbl.gov}}
({\em SNAP}) satellite in its primary mission 
of supernova photometry, and its additional capabilities for
weak lensing, photometric redshift, and astrometric surveys. 
In any astronomical camera design there is a trade-off in choosing an
angular scale for the pixels:  pixels small enough to finely sample
the instrument resolution will prevent the degradation or aliasing of
small-scale image information.  Larger pixels, however, ameliorate
read noise and may allow a
larger field of view (FOV) in cases where detector pixels or
bandwidth are scarce, or due to optics constraints.
How severe is the penalty in photometric accuracy that one
incurs from degraded sampling and resolution of larger pixels, and
when does the loss outweigh the potential gains in FOV?
Such tradeoffs are apparent in the 0\farcs1 pixel scale of the
WFPC2 wide-field CCDs.  Given the volume of data taken with WFPC2,
surprisingly few examinations of this tradeoff have been published.
Several recent publications in the astronomical literature have
discussed aspects of the more general exposure-time issues delineated
above: 
\begin{itemize}
\item \citet{L99a}[L99a] gives a good review of the mathematics of
undersampled images, and presents a method for removing aliased
signals given an arbitrary pattern of dithered exposures.  I will
follow the L99a conventions where possible.  
\item \citet{L99b}[L99b] continues with a
discussion of point-source photometry in undersampled images,
giving numerical results for the errors inherent in naive aperture
photometry (and centroiding) on {\em HST} images.  In this document I
discuss extensively what L99b briefly mentions:  that proper PSF-fitting
photometry can be much more accurate, both for flux and centroid, than
simple aperture-summing.
\item \citet{AK00} give a lengthy discussion of the derivation of
accurate astrometric information from WFPC2.  The PSF-fitting
techniques I use here would extract the same information from the
images. 
\item \citet{HF00} (and references therein) discuss the reconstruction
of dithered undersampled images, particularly the {\sc Drizzle}
algorithm, which is a robust spatial-domain technique.  In this
document I will be concerned not with image reconstruction, but with
quantitative extraction of various image moments (fluxes, centroids,
and ellipticities), so I'll not make use of {\sc Drizzle}.
\item \citet{KTL} describe several kinds of PSF figures of
merit---including several given in this paper---with regard to the
{\em WFHRI} concept of an array of ground-based 
tip-tilt telescopes.  Further information may be found on the web
pages for the {\em WFHRI} and {\em POI} projects.
\item Technical reports for space-based telescope projects
have addressed some of these issues, primarily with simulated data,
{\it e.g.} \citet{SHR99} investigate the undersampling issue for {\em WFC3},
\citet{RIL00} examine the maximum cosmic-ray load for {\em NGST}
instruments, and \citet{PS00} use the {\em NGST Mission
Simulator}\footnote{{\tt http://www.ngst.stsci.edu/nms/main}} to
investigate optimal pixel sizes.
\end{itemize}

All of the techniques used in this paper are familiar to the
image-analysis community and many of the elements are discussed in the
above and other references.  But I have not found in the astronomical
literature: an application of the Fisher information matrix to
point-source photometry in the presence of cosmic rays; a quantitative
discussion of the effects of pixel size on weak lensing measurements;
or a quantitative derivation of the required amount of dithering.
More importantly, there is not to my knowledge a 
publication or software tool which combines all of these
important effects to make detailed exposure-time
estimates. That is the goal of this publication.

Following this Introduction is a general discussion of pixelization
and sampling upon imaging observations, giving the analytical
framework for the calculations.  The next section briefly describes the
implementation of these ideas in the {\tt ETC++} software package.
\S\ref{results} demonstrates the capabilities of the methods and
software by providing quantitative answers to some general questions:
what is the $S/N$ penalty for oversized pixels?  What amount of
dithering is required to reach optimal $S/N$?  Then I address some
more specific questions about optimizing camera configurations, and
comparing the performance of state-of-the art space-based imaging vs
ground-based imaging.

\section{Pixelization, Sampling, and Noise}

\subsection{Fourier Description}
Following the L99a and L99b exposition:  The scene being imaged has
intrinsic intensity distribution $O(x,y)$, with Fourier transform
$\tilde O(k_x, k_y)$.  The Fourier transform convention is ``System
2'' of \citet{Br78}:
\begin{eqnarray}
O({\bf x}) & = & (2\pi)^{-2} 
	\int d^2\!k\, \tilde O({\bf k}) e^{-i{\bf k\cdot x}} \\
\tilde O({\bf k}) & = & \int d^2\!x\,O({\bf x}) e^{i{\bf
k\cdot x}}.
\end{eqnarray}
The telescope optics convolve the image with some optical point-spread
function (PSF) $P(x,y)$ (which I take to have unit integral).  With the
above convention for the Fourier transform, the convolution $O\ast P$
has transform $\tilde O \cdot \tilde P$.

The pixelization of the image by the detector entails two operations:
first, the optical image is convolved with the {\bf pixel response
function} (PRF) $R(x,y)$ (which I normalize to unit integral), and sampled
on the two-dimensional grid of pixel centers on spacing $a$.  The data
from a single array readout are thus the image
\begin{equation}
I(x,y)=\left[O(x,y)\ast P(x,y) \ast R(x,y) \right] 
	\shah({x\over a},{y \over a}) 
\end{equation}
where $\shah$ is the 2d {\em shah} function,
\begin{equation}
\shah(u,v)\equiv
\sum_{i=-\infty}^{+\infty}
\sum_{j=-\infty}^{+\infty}\delta\left(u-i\right)
\delta\left(v-j\right).
\end{equation}
In the Fourier domain, the pixelated, sampled image is
\begin{eqnarray}
\tilde I({\bf k}) & = & [\tilde O({\bf k}) \tilde P({\bf k}) \tilde R({\bf
k})] \ast a\shah\left( {a k_x \over 2\pi}, {a k_y \over 2\pi}\right) \\
 & = & \sum_{m=-\infty}^\infty \sum_{n=-\infty}^\infty
	\tilde O \tilde P \tilde R(k_x+m\Delta k, k_y+n\Delta k) \\
\Delta k & \equiv & {2\pi \over a}.
\end{eqnarray}
The detected image, therefore, looks like the source image as
convolved with an {\em effective PSF} (ePSF) $P^\prime\equiv P\ast R$, and
sampled at interval $a$.  The sampling mixes power at
spatial frequency $k_x+m\Delta k$ down to frequency $k_x$, leaving the
nature of the original $\tilde O({\bf k})$ ambiguous.  This {\em
aliasing} is detrimental to our efforts, as we cannot from a single
measurement know exactly
either the ePSF (from observing point-source stars) or the intrinsic
scene $O$.

An optical
telescope cannot transmit spatial frequencies beyond
$\pm k_{\rm max}=\pm 2\pi
D / \lambda$, where $D$ is the largest dimension of the telescope
aperture and $\lambda$ is the wavelength.  There is no aliasing if
\begin{equation}
\Delta k \ge 2 k_{\rm max} \quad \Rightarrow \quad 
a \le \lambda / 2D.
\end{equation}
For $D=2$~meters, $\lambda = 1\,\mu{\rm m}$, this {\em Nyquist}
sampling corresponds to 0\farcs05 pixels.

When the data have been sampled at Nyquist or higher density, we can
produce shifted, rotated, or deconvolved versions of the image with no
ambiguity (apart from noise).  In the {\em SNAP} mission, this will mean
that subtraction of the host galaxy from supernova images will be
essentially perfect, as long as the template image is
Nyquist-sampled.  This holds for other time-domain signals, such as
microlensing, planetary transits, and Kuiper Belt surveys.  For weak
lensing surveys, it means that the systematic ellipticities imposed on
galaxies by the PSF can, in theory, be removed nearly perfectly.
Nyquist sampling is thus highly desirable.

\subsection{Dithering}
By taking a series of exposures with
pointings {\em dithered} by a fractional pixel amounts, we can sample
the ePSF-convolved scene more densely than the pixel grid.  

It is important to realize that the two effects of pixelization are in
fact separable:  the ePSF depends on the size of the pixel through the
PRF $R(x,y)$; but {\em the sampling density can be denser than the
pixel grid $a$}.  If we choose dither positions on a grid $a/N$, then
we eliminate aliasing as long as $k_{\rm max} < N\pi / a$.  We can
therefore obtain Nyquist-sampled data even with large pixels.  To
first order this comes with no noise penalty:  if we replace a single
exposure of time $T$ with a dithered grid of $N^2$ exposures each of
time $T/N^2$, then the total counts from the source are the same; the
final image has the same number of sky photons per unit area (fewer
per sample, but more samples per unit area).  There is, however, an
increase in overhead and read noise from the extra exposures, and the
data rate must be higher.

What is the optimal dither pattern?  L99a demonstrates that, for image
reconstruction, a regularly interlaced grid offers the lowest noise.  I have
not encountered any reason to execute any other pattern.  Interlacing
makes the analysis straightforward, and the L99a and {\sc Drizzle}
techniques can be rendered equivalent in this case.

Given that interlacing can recover Nyquist sampling, the remaining
drawback to larger pixels is the poorer resolution in the ePSF, which
degrades the S/N for background-limited photometry and for centroid
and ellipticity measurements of marginally resolved galaxies.  I will
quantify this below.

\subsection{Space vs Ground}
There are two important differences between space-based and
ground-based data---one obvious and one more subtle---that
suggest that a spaced-based observatory is likely to
make use of larger pixels (relative to the PSF FWHM) 
than a ground-based imager:
\begin{enumerate}
\item {\bf Dithering does not work easily for ground-based images.}  This is
because atmospheric seeing is constantly changing the PSF.  Each
successive exposure would be sampling a different PSF, rendering the
de-aliasing difficult or impossible.  If the PSF varies on spatial
scales of $\Delta\theta$, then there must be enough PSF-template stars
in each $\Delta\theta^2$ area to sample different pixel phases and
solve for the unaliased PSF.  Space
observatories, however, can have exceptionally stable PSFs, longer
than the time required to complete a dither sequence.  For {\em HST}, the
PSF varies significantly on the 90-minute orbit time due to thermal
cycling.  {\em SNAP} will rarely go into Earth eclipse, and even the
Earthshine thermal load will vary only on the 15-day orbital period.

\item {\bf The ``dynamic range'' of a ground-based PSF is larger
than in space.}  Colloquially an image is considered
``Nyquist-sampled'' if there are $\gtrsim2$ pixels across the FWHM of
the PSF.  On large (8-meter) ground-based telescopes in excellent
seeing, this requires pixels $\lesssim 0\farcs2$, which is easily
accommodated, in fact difficult to avoid given the plate scales of 8m
telescopes.  In 
fact the PSF can have structure all the way to $\lambda/2D$, which is
only 6~mas for $V$-band observations, so a formally complete sampling
of the PSF is not practical.  The high-$k$ image power is strongly
suppressed by the atmospheric seeing, however, so there is some
sampling density at which the aliased power can be deemed
insignificant.  For a PSF generated by Kolmogorov turbulence
($\tilde P=\exp[-(k/k_0)^{5/3}]$), sampling at 2.5 pixels per FWHM
limits the aliased Fourier amplitude to about 1\% of the total
amplitude.  This may not suffice for some applications, such
as high-precision difference imaging, or weak-lensing surveys which
need systematic ellipticity errors reduced to $\sim10^{-4}$.  Cutting
the aliased amplitudes down to 0.1\% requires 3.1 samples per FWHM.

A diffraction-limited circular telescope, on the other hand, has a FWHM of
$\approx 1.0\lambda/D$ and can have no structure shorter than
$0.5\lambda/D$.  Hence putting $\approx2$ pixels (or samples) across
the FWHM leaves {\em no} ambiguities.
\end{enumerate}

The strict cutoff of the Airy PSF at $k=2D/\lambda$ also means that
space-borne observatories will be relatively insensitive to pixel
response functions that depart from the ideal unit-square model.  If
the PRF has structure at wavelengths $\lesssim 1/5$ the pixel spacing,
it will be irrelevant, since the PSF does
not pass spatial frequencies much smaller than the FWHM, which will be
close to the pixel size.  Similarly, subtle pixel-to-pixel variations
in the PRF will not matter if they occur at high spatial frequencies.
In \S\ref{results} we will quantify the effect of a sharply bounded
``dead zone'' within each pixel, which can be taken as an extreme case
of intrapixel variation.

All space-based observations intended for use as image-differencing
templates or weak-lensing measurements should be interlaced by a
factor $N$ sufficient to reach Nyquist sampling.  It is {\em not}
necessary to Nyquist-sample exposure sequences intended solely for
photometry of time-variable point sources, as investigated below.

\subsection{Fisher Information for Point-Source Photometry}
The most accurate method for point-source photometry is PSF-fitting.
The vector of unknown parameters ${\bf p}=\{f,x_0,y_0\}$ (fluence and
position) is varied to minimize deviations from a model PSF.  We minimize
\begin{equation}
\label{chisq}
\chi^2 = \sum_i { 
\left[ \hat I({\bf x}_i) - f P^\prime({\bf x}_i-{\bf x}_0) \right]^2
\over {\rm Var}(I({\bf x}_i)) },
\end{equation}
where $\hat I$ is the measured fluence (counts per readout) at
position ${\bf x}_i$ and $P^\prime$ is the ePSF.  Minimization of
$\chi^2$ is equivalent to maximizing the likelihood of the
observations under the assumption of Gaussian noise statistics.
The {\em Fisher information matrix} $F$ is defined as the second
derivative of the (log) likelihood function with respect to the
parameters.  In our case this becomes
\begin{equation}
\label{fisher}
F_{ij} = \sum_{\rm pixels} {{ \partial (fP^\prime) \over \partial p_i}
	{ \partial (fP^\prime) \over \partial p_j}  \over
	{\rm Var}(I({\bf x}_i)) }.
\end{equation}
It can be shown that the inverse of the Fisher matrix is the best
attainable covariance matrix for the unbiased parameter estimates (if
a best indeed exists) \citep{KS69}.  Fisher matrices have been widely
discussed and applied to cosmic background anisotropy measurements,
for example by \citet{TTH97}.

If the centroid is known {\it a priori}, then the uncertainty in fluence
is just ${\rm Var}(f) = (F_{ff})^{-1}$.  To calculate the Fisher
matrix, we need the ePSF, the incident flux, and a noise model.  
The sum runs over all pixels in all exposures of a sequence.  
I will describe an observing sequence by the interlacing factor $N$,
and by the number $M$ of exposures taken at each of the $N^2$ dither
positions.  In {\em HST} parlance, $M$ is the ``CR-split.''

Ideally the noise is dominated by shot noise from the source and from
a uniform sky background of $n$ counts per unit solid angle per
readout.  In this 
case ${\rm Var}(I) = fP^\prime + na^2.$ In the limit of bright
sources, the flux uncertainty reduces to ${\rm Var}(f)/f^2 =
(fMN^2)^{-1} = N_\ast^{-1}$, ($N_\ast$ is total source counts),
independent of the pixel or dithering configuration.  

In the
background-dominated limit, when we take the centroid as fixed, the
flux uncertainty simplifies to
\begin{equation}
(S/N)^{-2} = { {\rm Var}(f) \over f^2} = { na^2 \over f^2 
	\sum [P^\prime({\bf x}_i)]^2}.
\end{equation}
If the image has been sampled at the Nyquist density or higher, then
we can apply Parseval's Theorem to obtain the simple form
\begin{eqnarray}
\label{AreaSN}
(S/N)^{2} & = &  { N_\ast^2 \over nNM^2\, A_{SN}} \\
A_{SN} & \equiv & 4\pi^2\int d^2\!k\, |\tilde P^\prime({\bf k})|^2 \\
 & = & 4\pi^2\int d^2\!k\, | \tilde P^2({\bf k})|\,| \tilde R^2({\bf k})|.
\end{eqnarray}
Thus in the limit of faint, Nyquist-sampled, unresolved sources, the
$S/N$ for detection/photometry depends up the effective area $A_{SN}$
of the PSF, and we can easily see how the PRF affects this.  The
$k$-space integral form is particularly convenient since the Airy PSF
is bounded to $k<2D/\lambda$, and no convolutions must be executed.
\citet{KTL}, for example, derive and make use of this form.

Read noise and dark current produce white noise that can be subsumed
into $n$ in the simple formula (\ref{AreaSN}).  When the image is not
Nyquist sampled, it is easier to transform the ePSF to $x$-space 
and use \eqq{fisher}.

The Fisher matrix also allows us to evaluate the astrometric accuracy
for point sources.  This is not a primary goal for {\em SNAP}, but I will
present some results in \S\ref{astrometry}.  In the background-limited
case, the one-dimensional uncertainty of point-source astrometry is
simply quantified as
\begin{eqnarray}
\label{AreaCentroid}
\sigma_x^{2} & = &  { nNM^2\, A^2_{\rm cent} \over N_\ast^2 }\\
A^2_{\rm cent} & \equiv & 4\pi^2\int d^2\!k\, |k_x\tilde P^\prime({\bf
k})|^2 \\ 
 & = & 4\pi^2\int d^2\!k\, k_x^2\, | \tilde P^2({\bf k})|\,| \tilde R^2({\bf k})|.
\end{eqnarray}
This and related forms are given by \citet{KTL} and used for analysis
of the proposed {\em POI} project.

\subsection{Cosmic Rays}
The incorporation of cosmic-ray (CR) hits into the Fisher formalism is
easy.  We just remove from the sum (\ref{fisher}) the information
contributed by pixels that are ruined.
In the {\em SNAP} mission, we expect the CRs to span many pixels,
while the ePSF will be $\lesssim 2$ pixels across.  Hence the probability of
losing the entire exposure's information is essentially equal to the
probability $P_{CR}$ of the central pixel being contaminated during an
exposure.  For detectors with non-destructive readout (such as HgCdTe
arrays), being sampled continuously during the exposure, the
information lost is only that fraction accumulated {\em after} the CR
hit. 

There should be little difficulty identifying cosmic rays in space-borne
images, as the vast majority of hits cover many pixels and deposit
thousands of electrons.  On the ground, cosmic rays cause negligible
loss of information.

\subsection{Galaxy Photometry}
\label{galphot}
In the {\em SNAP} mission it will be important to derive accurate colors for
resolved galaxies so as to obtain photometric redshift estimates for
host galaxies.  This places performance requirements on the $S/N$ of
galaxy photometry.
For a galaxy with known intrinsic flux distribution $g({\bf x})$, the
best possible $S/N$ on the total flux is derivable through the same
Fisher information formalism as for point sources.  This is equivalent
to measuring the flux in a Wiener-filtered image.
In practice,
however, galaxies come in an infinite variety of shapes, so one cannot
{\it a priori} choose the ideal filter for each image.  More practical is
to measure the flux through some predetermined aperture of shape
$w({\bf x})$:
\begin{equation}
\label{fw}
f_w \equiv \int d^2x\,  w({\bf x}) I({\bf x}).
\end{equation}
This weighted flux is not useful for studies requiring absolute total
luminosities for galaxies, but will provide very accurate galaxy
colors if matched apertures are used in different wavelength bands.
I will assume, for simplicity, that both the galaxy and the weight
are circularly symmetric.  Then a simple propagation of errors gives a
$S/N$ ratio for the $f_w$, when Nyquist-sampled, of
\begin{equation}
\label{galsn}
\left({S\over N}\right)^{-2} = 
(2\pi)^2 { \int d^2k\, \tilde g \tilde P^\prime \widetilde{w^2} \, + \,
	n \int d^2k\, |\tilde w^2| \over
	\left[\int d^2k\, \tilde g \tilde P^\prime \tilde w \right]^2}.
\end{equation}
The first term in the numerator is the source shot noise, the second
term is from the white-noise background arising from sky (or dark,
read) counts of $n$ per unit area.

A useful choice of weight is the Gaussian, $w=e^{-r^2/2\sigma^2}$.
The Fourier transform is of course also a Gaussian, and \eqq{galsn}
can be evaluated for any candidate ePSF $P^\prime$ and galaxy profile $g(r)$.
The size of the weight function $\sigma$ can be
adjusted to optimize the $S/N$ for each galaxy on an image.

Clearly the effect of finite resolution in the PSF or PRF is to remove
high-$k$ information which might be present in the galaxy image.
If the galaxy scale is larger than the PSF, then the PSF is irrelevant.
Small galaxies reduce to the point-source limit.

Below I evaluate the resultant $S/N$ for exponential-disk galaxies
($g\propto e^{-\alpha r}$).  In this case a Gaussian-weighted flux
measurement is only a few percent noisier than the optimally-weighted
measurement.  

\subsection{Galaxy Ellipticities}
For weak gravitational lensing measurements, we wish to detect small
shears to the intrinsic shapes of galaxies.  A poorly resolved
or noisy galaxy image will inhibit this.  It can be shown \citep{BJ01}
that for a background-limited, nearly
circular, Nyquist-sampled galaxy with radial flux profile $g$, the
photon-noise contributions to the uncertainty in the ellipticity
components $e_1$ and $e_2$ are each optimally
\begin{equation}
\label{evar}
\sigma^2_e = {32 \pi^2 n \over f^2}
	\left[ \int d^2k\, \left| k \tilde P^\prime({\bf k})
	{ \partial \tilde g \over \partial |k| } \right| ^2
	\right]^{-1}.
\end{equation}
This is again easily calculated if we assume an exponential $g$ and
know our ePSF.  In weak-lensing measurements, this photon noise level
must be reduced below the {\em shape noise} level of $\sigma_e\approx 0.3$
attributable to the intrinsic variation in galaxy ellipticities.
Non-circular galaxies will have slightly more photon noise than
\eqq{evar} for a given magnitude and size, but this is a minor effect.

A further limitation to weak lensing measurements is systematic
contamination of the intrinsic shapes by uncorrected artifacts of
asymmetric ePSF's.  As noted above, a Nyquist-sampled space-telescope
image provides complete information on the ePSF, and hence will permit
nearly-perfect suppression of these systematic errors.  
Ground-based images must be sufficiently well-sampled to avoid
aliasing any significant power.  Furthermore, as the PSF is temporally
and spatially variable in ground-based images, the mean spacing 
between bright
PSF-template stars must be less than the angular scale of PSF
variation. The space telescope has the luxury of constructing a PSF 
map by combining template stars from a series of exposures.

\section{C++ Implementation}
The above formulae have been implemented as a set of {\tt C++} classes
and driver programs.  This {\tt ETC++} software is available from the
author.  The interesting elements of the code are described here.

\subsection{Classes}

\subsubsection{ {\tt Psf}}
PSFs or ePSFs can be created as instances of the {\tt Psf} class.  A
{\tt Psf} can be an Airy pattern, a Gaussian, a square PRF, a
Kolmogorov-turbulence seeing function, or an arbitrary  convolution of
any of these.  Any {\tt Psf} can return its value at some ${\bf k}$
vector, or the real-space transform; the point-source sensitivities
$A_{SN}$ and $A^2_{\rm cent}$ can be calculated; or, given a {\tt
Galaxy} specification, the photometric accuracy \eqq{galsn} or ellipticity
variance \eqq{evar} can be calculated.

\subsubsection{ {\tt Galaxy}}
The {\tt Galaxy} base class describes a galaxy image.  A {\tt Galaxy}
has a flux and half-light radius.  One can request the intensity of the
galaxy at any point in $x$ or $k$ space, or the derivative $dg/d(\ln
k)$ required for \eqq{evar}.  There are currently two options:
{\tt GalGaussian} and {\tt GalExp}.

\subsubsection{ {\tt Params}}
The {\tt Params} class contains all the specifications of an
observatory and observing scheme that are necessary to calculate the
$S/N$ quantities:  telescope aperture, obscuration, quantum
efficiencies, filter specifications, detector characteristics,
exposure times and sequences, cosmic-ray rates, etc.  These are input
from text files, and then can be parsed to produce the effective {\tt
Psf} for the observation, the source and sky count rates, etc.

\subsubsection{ {\tt Fisher} Tools}
Given an observing scheme, count rates, detector noise model, and
ePSF, the {\tt FisherCalc} class can produce the Fisher information
matrix using \eqq{fisher}, then report the parameter errors.
If cosmic rays are present, then there is a distribution of possible
flux and centroid errors:  this error distribution is calculated
either by an exhaustive search of possible cosmic-ray outcomes, or by
Monte-Carlo sampling of cosmic-ray outcomes.  The uncertainty
distribution can also include an integration over a grid of possible
source positions relative to the pixel grid.

\subsection{Executable Programs}
There are top-level programs that make use of the above classes to 
return the photometric or astrometric $S/N$ of point sources for a
given observing scheme.  Because the $S/N$ depends upon pixel phase
(for undersampled images) and cosmic-ray outcomes, the $S/N$ levels
are in fact reported as percentile values, {\it e.g.}
the median, 5th and 95th percentile flux errors.
Other top-level programs report the photometric and ellipticity
measurement speeds for galaxies.

More interesting is the {\tt optimize} program, which
seeks the observing scheme (exposure
times, interlacing and CR-split factors) that reaches a desired $S/N$
on point sources in minimum total observing time.  The target $S/N$
must be specified, as well as the ``confidence level'' giving the
fraction of sources which must be measured to the target accuracy.
This calculation is meant to be quite realistic, including all
time overheads, cosmic rays, sampling, etc.   

\section{Applications}
\label{results}
This section demonstrates some applications of the above tools.  Some
of the questions addressed here are very general, while some are
specific to the {\em SNAP} optimization.

\subsection{How Much Dithering Is Required?}
If the pixels are larger than the Nyquist size $0.5\lambda/D$, what
interlacing factor $N$ is required to recover most of the
photometrically useful information?  Estimation of the point-source
flux via PSF-fitting does not require Nyquist sampling, as long as an
unaliased PSF template is available; the loss of
information from aliasing can be minor if the sampling is adequate.

\begin{figure}[t]
\plottwo{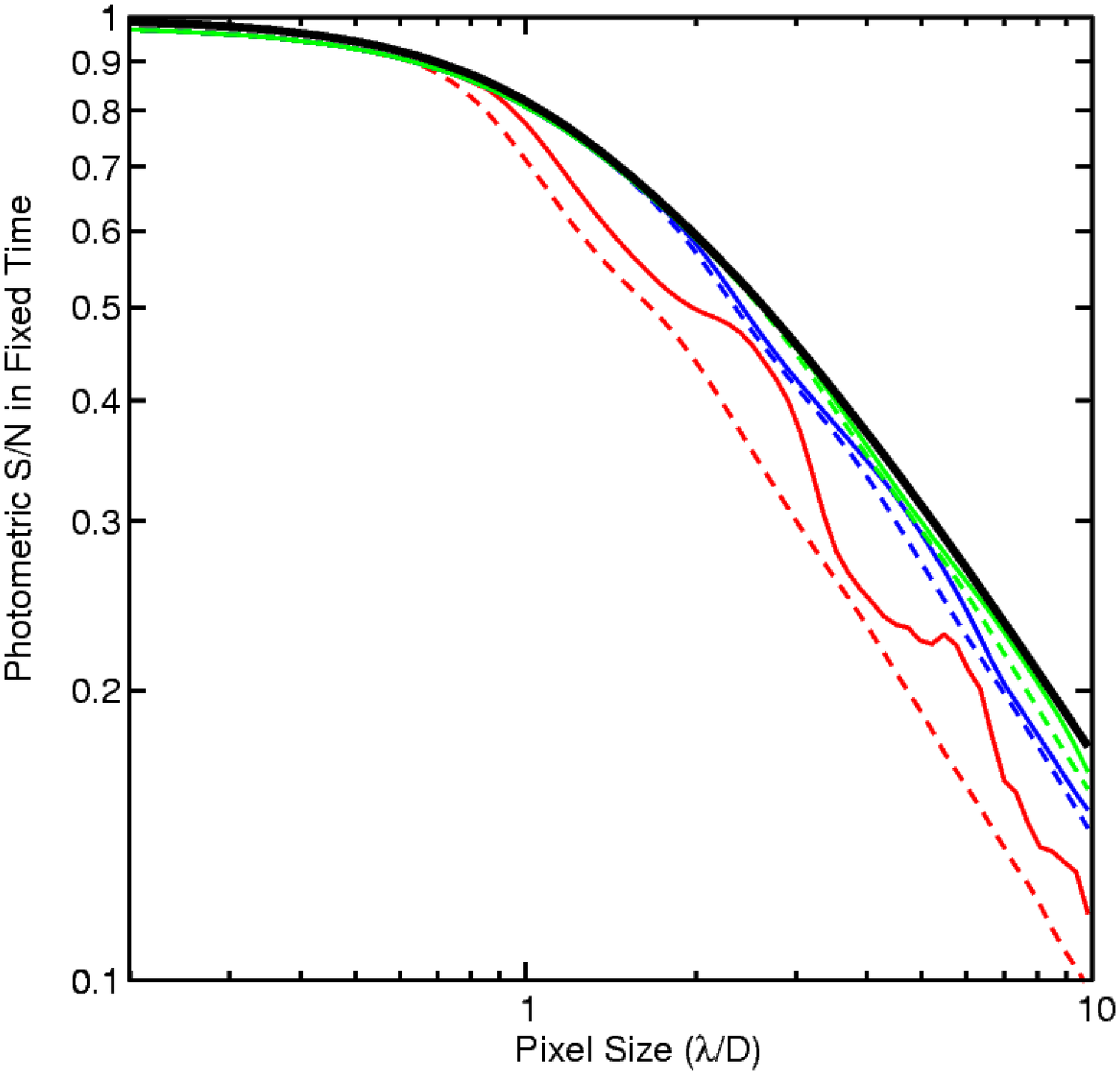}{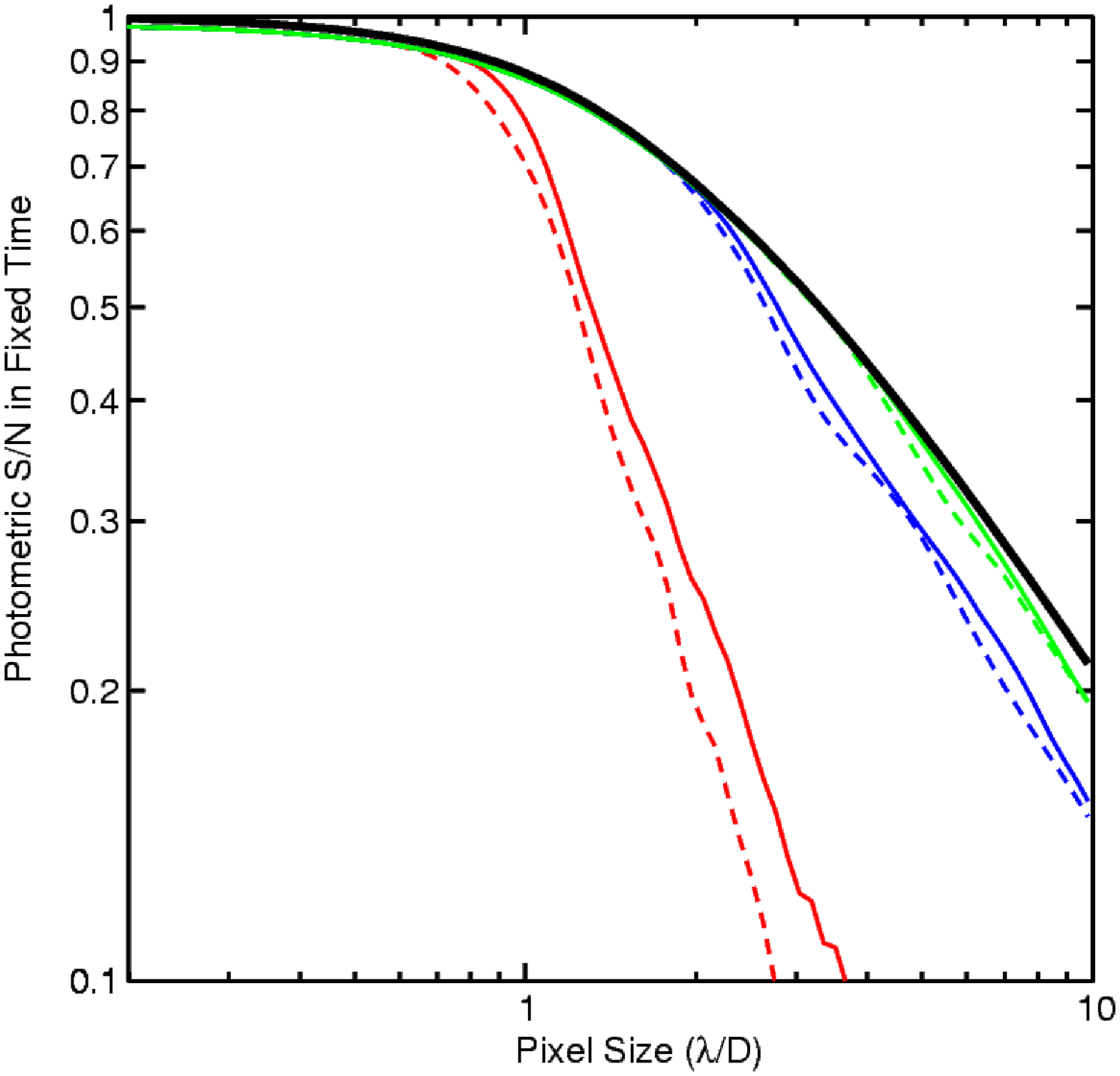}
\caption[]{
\small The $S/N$ ratio attained in fixed integration time is plotted as a
function of the pixel size (in units of $\lambda/D$) for the limiting
case of background-limited, uncrowded point-source photometry with no
readout overheads and a diffraction-limited optical PSF.  In the
left-hand panel, the pixel response 
function is a perfect square, but on the right-hand side the pixels
are assumed to have a dead zone in the outer 10\% of each edge.
The heavy black line assumes that the image has
been interlaced to reach Nyquist sampling.  The solid red line is the
median $S/N$ when there is no interlacing at all; the $S/N$ at the
least favorable pixel phase is the dashed red line.  The blue and
green lines give the corresponding data for exposure interlacing
factors of $N=2$ and 3, respectively.  Interlacing by a factor 3
recovers nearly all the available $S/N$ in every case.
}
\label{chooseN}
\end{figure}

Figure~\ref{chooseN} shows the recovered $S/N$ ratio for a
diffraction-limited PSF observed on 
perfect square pixels of some size $P\lambda/D$,
for various interlacing factors $N$.  The total observing time is held
fixed, I ignore overheads and read noise, and assume here a
background-limited observation.  The heavy line shows the $S/N$
for Nyquist-sampled images (relative to infinitesimal pixels). 
The degradation of point-source $S/N$ as the PRF broadens is apparent.
The solid red line shows the median $S/N$ in the $N=1$ case, {\it i.e.}
no interlacing.  There is up to 30\% degradation when $P>1$.  The dashed
red line shows the $S/N$ at the worst pixel phase.

A little dithering helps a lot, however.  For $N=2$ (blue lines),
both the RMS and worst-case are very close to Nyquist for
$P\lesssim8$. 
For $N=3$ (blue lines), both RMS and worst-case are within a few percent
of Nyquist at all pixel sizes.  Photometrically speaking, therefore,
there is little point to interlacing at $N>3$.  The reason is that the
PRF itself rolls off sufficiently quickly that there is little aliasing
for $N>3$, regardless of the PSF.

Many detectors do not have uniform response across the geometric pixel
square.  As a canonical example I consider a case where the pixel contains
dead ``gutters'' at the outer 10\% of each pixel edge, so the active
area covers only 64\% of the geometric pixel.  For oversized pixels
one might worry that the star could fall into the dead area.  Since
the PSF is not finite, there is always some flux in the sensitive
areas.  While naive aperture photometry will fail in this case, 
a PSF fit will recover an unbiased flux estimate and centroid---but 
the loss of information could be significant, as intuition suggests.
The second panel of Figure~\ref{chooseN} shows the photometric $S/N$
vs pixel size in this case.  Non-interlaced exposures ($N=1$, red
line) are far worse than Nyquist sampling for $P>1$.  Interlacing with
$N=2$ recovers the Nyquist $S/N$ up to $P\approx2$, however, and $N=3$
interlacing again recovers almost the full Nyquist $S/N$ for any
sensible pixel size.

Once the ePSF is known from bright stars, the non-uniformity
of the PRF is immaterial.  Of course if each pixel has a different PRF,
then the template ePSFs will be incorrect, leading to magnitude
errors.  But it is important to recall that only variations at spatial
frequencies below $2\pi D/\lambda$ can make any difference.  Unless
the PRF is grossly larger than the Airy disk (and the charge-diffusion
scale), inter-pixel variations will be strongly damped in the ePSF.

\begin{figure}[t]
\plottwo{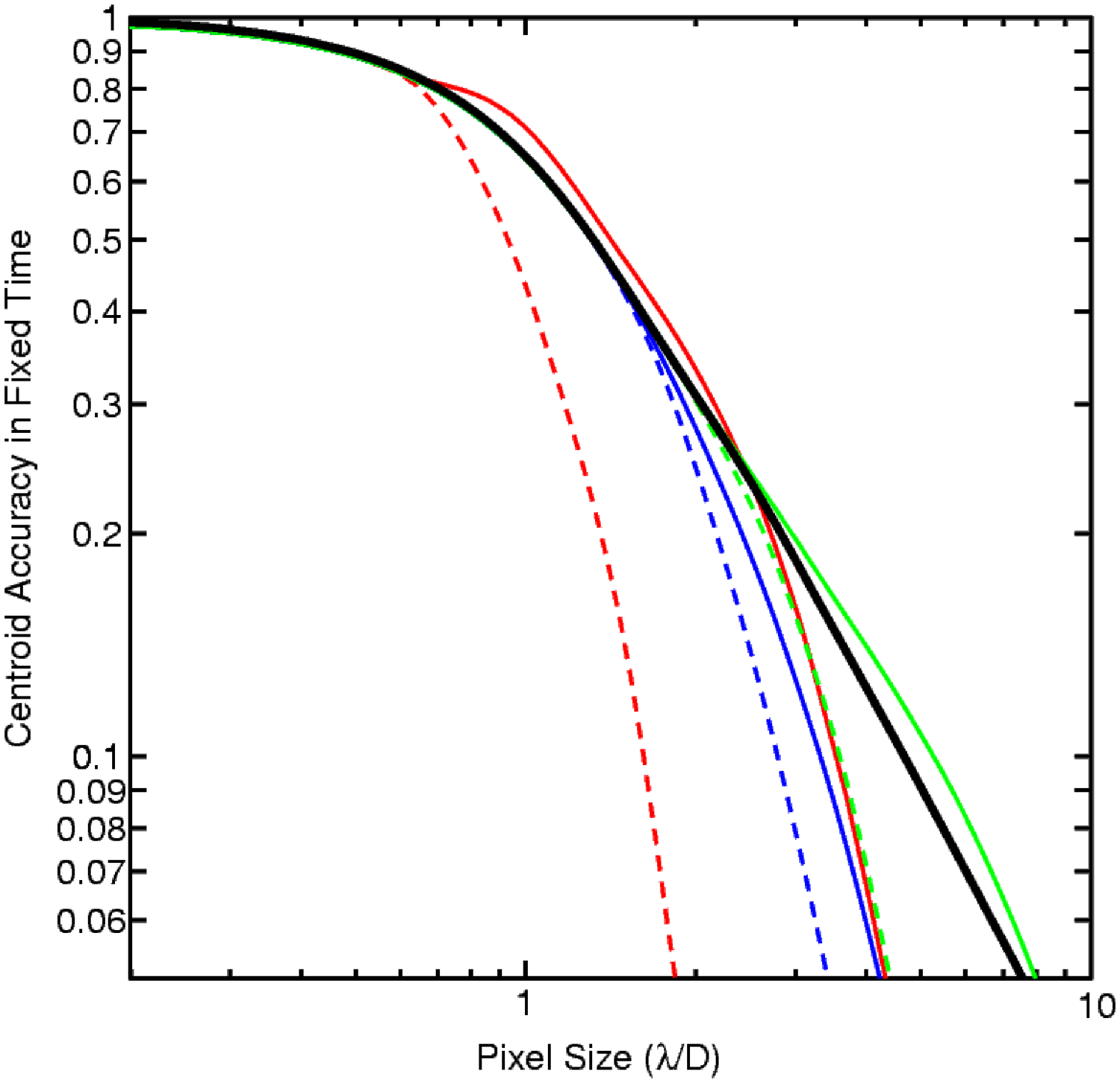}{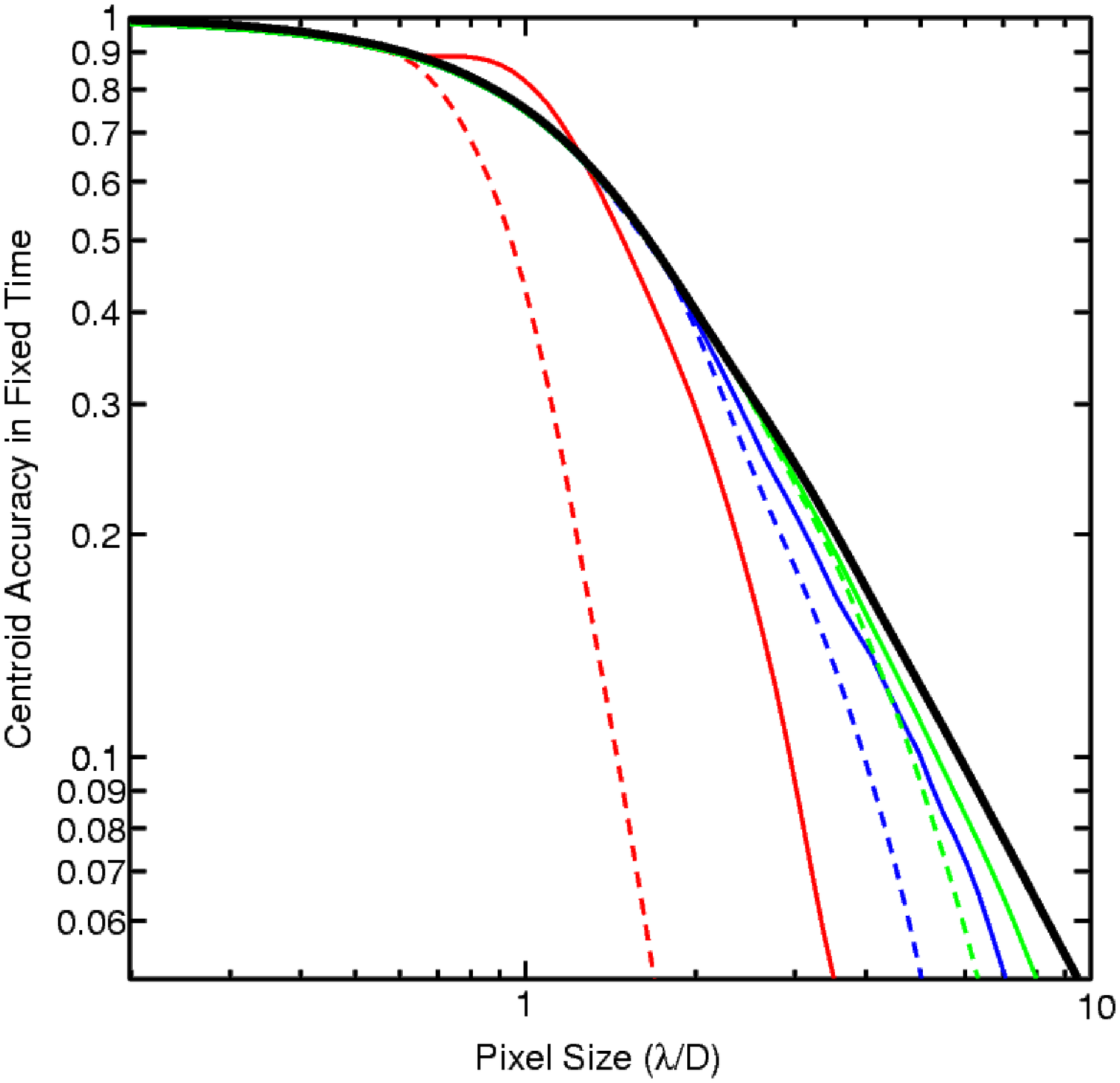}
\caption[]{
\small The accuracy of the centroid determination for a
background-dominated point source in fixed
observing time is plotted as a function of pixel size.  
As in Figure~\ref{chooseN}, the left-hand side is for an ideal square
pixel and the right-hand side is for pixels with a 10\% ``gutter.''
Line types are also as in  Figure~\ref{chooseN}.
The penalty for large pixels is more severe for astrometric
observations than for photometry.  For large pixels we see that $N=3$
interlacing is, in 
the median case, nearly as good (or better!) than Nyquist interlacing,
but if the star falls at an unfavorable pixel phase, the astrometric
accuracy can be greatly degraded when $P\gtrsim2$.
}
\label{centrNyq}
\end{figure}

I plot in Figure~\ref{centrNyq} the relative {\em position} accuracy
for a pixellated Airy PSF as a function of pixel size and
interlacing.  
The penalty for large pixels is more severe for astrometric
observations than for photometry.  Interlacing at $N=2$ approaches
Nyquist centroiding errors for $P\lesssim2$ in all cases; likewise
$N=3$ recovers all information up to $P\lesssim3$.  For $P>3$ we
see that $N=3$ interlacing recovers the Nyquist accuracy in the median
case, but an unfavorably positioned star can have greatly degraded
astrometric accuracy.

In contrast to the photometric measurement, an astrometric measurement
of a bright star is degraded by large pixels.  But the dependence of
$\sigma_x$ on $P$ is not as steep as in the faint (background-limited)
case. 

For point-source measurements, therefore, I find it is typically
necessary to interlace exposures only to about half the Nyquist
density, though complications may arise for very large ($>3\lambda/D$)
pixels. If a large number of exposures must be taken, however---{\it
e.g.} to avoid saturation or cosmic-ray loading---one might as well
interlace to the Nyquist level.

\subsection{Optimization of Pixel Scale}
What choice of pixel scale allows a science goal to be achieved with
the fewest resources?  The most typical scarce resource is total
observing time $T$.  There may be an additional constraint on the FOV
imposed by the optical design, in which case one typically tries to
reduce the pixel scale $P$ until read noise is
important or $P\lesssim0.5$.  Less obvious is the optimal
pixel scale $P$ in cases where the number of pixels $N_{\rm pix}$ has
an upper bound imposed by detector cost, telemetry bandwidth
limitations, or engineering constraints.  

\subsubsection{Background-limited Point-Source Photometry}
\begin{figure}[t]
\plottwo{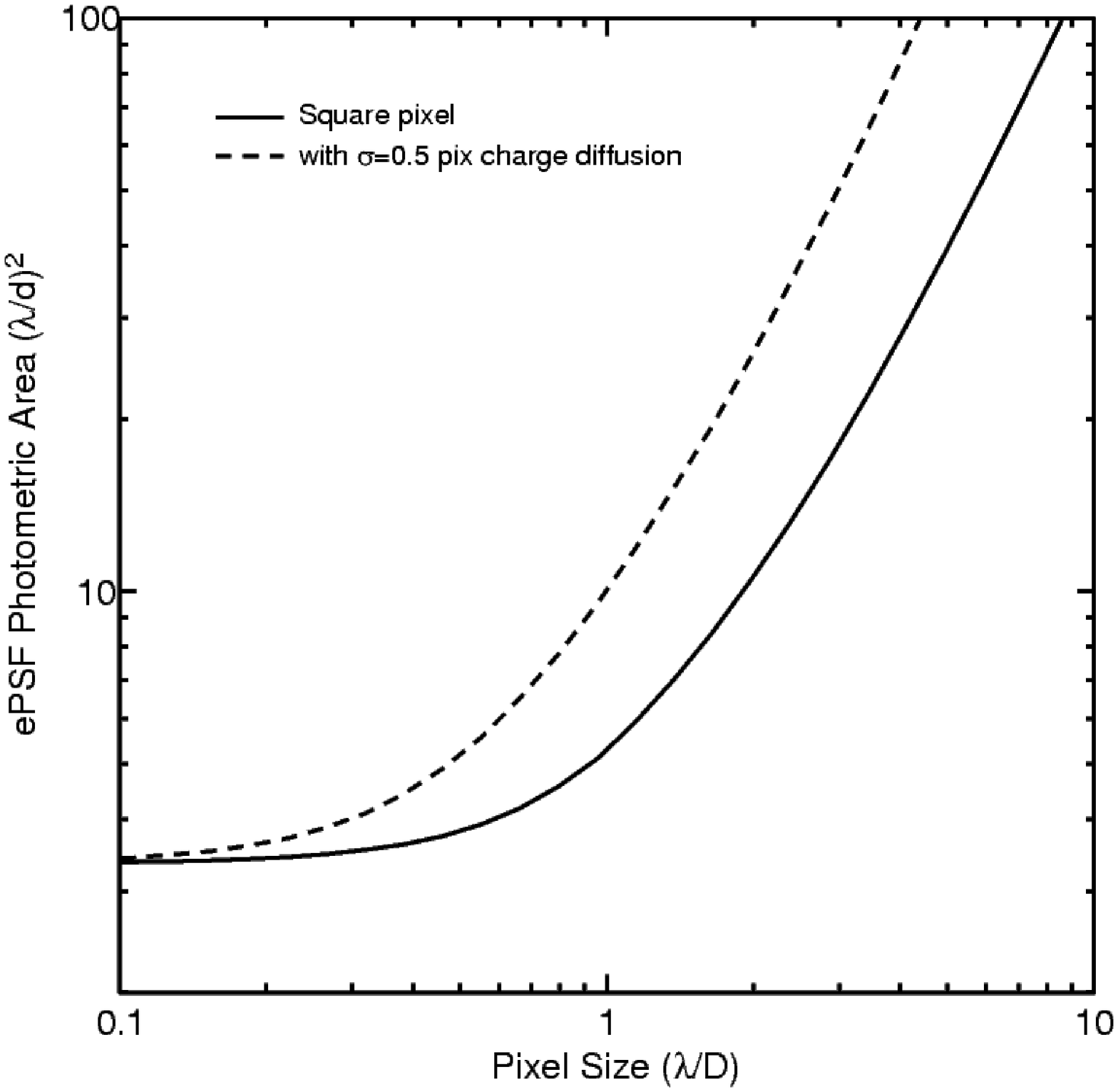}{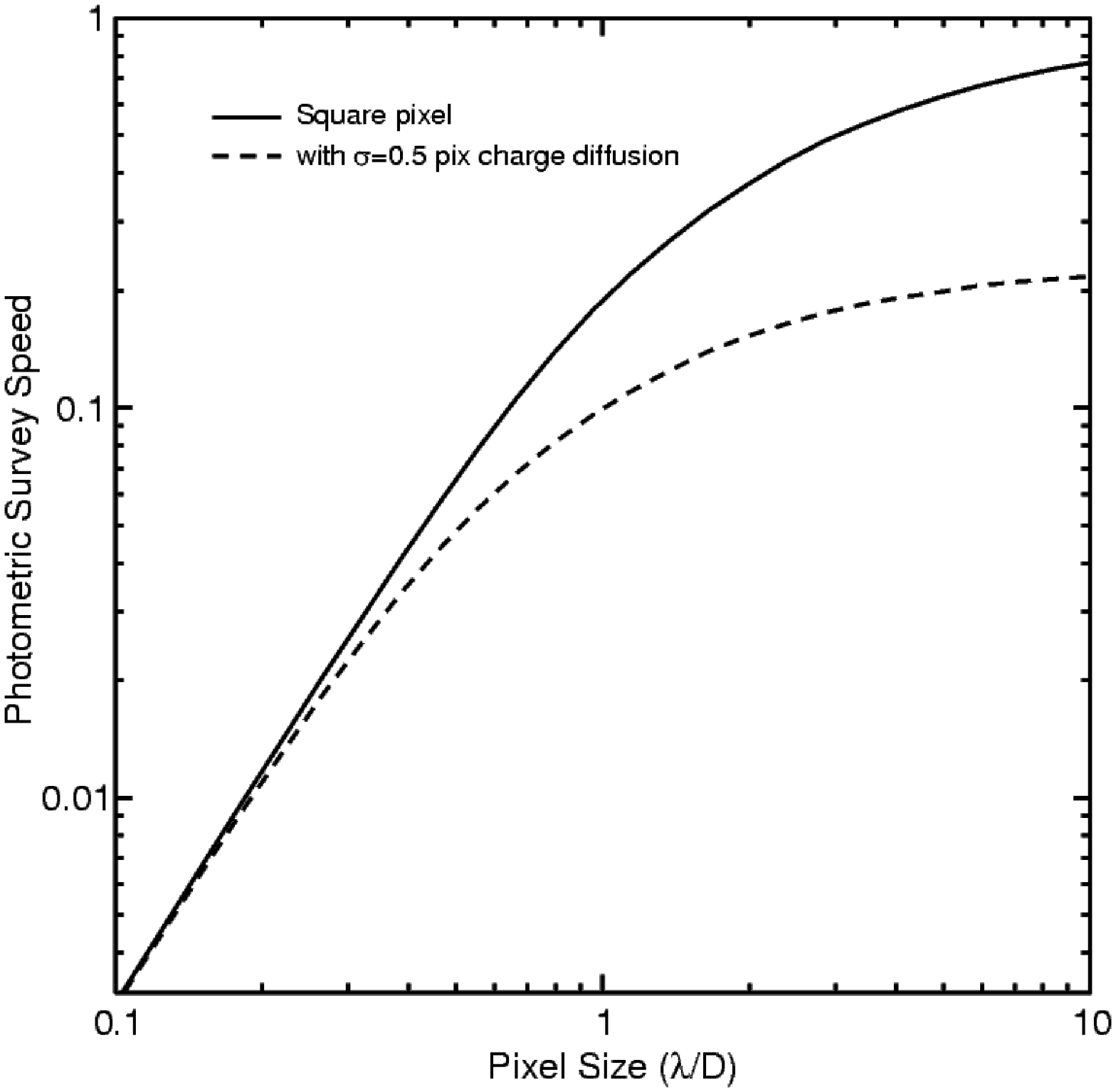}
\caption[]{
\small
On the left plot, the effective area $A_{SN}$ of the ePSF for
background-limited point-source 
photometry is plotted vs pixel size.  The solid line is for a perfect
square pixel atop a diffraction-limited PSF from a telescope with 30\%
pupil obscuration.  The dashed line shows the effect of charge
diffusion with $\sigma=0.5$~pixels.  The time to complete an
observation to fixed $S/N$ will scale as this parameter.  The penalty
from charge diffusion is substantial.  On the right side is the speed
to survey a given sky area to a given $S/N$, assuming that the number
of pixels is fixed.  Larger pixels always help, as long as the FOV
increases proportionately.
}
\label{ASNfig}
\end{figure}

The tradeoffs are most easily understood for background-limited
point-source photometry, in which case the detector geometry is fully
described by the $A_{SN}$ ``PSF area'' in \eqq{AreaSN}.  Consider the
PRF to be a square of angular size $P\lambda/D$.  Figure~\ref{ASNfig}
plots the value of $A_{SN}$ as a function of $P$.  In the limit $P\ll1$
of fine sampling, $A_{SN}$ reduces to that of the Airy pattern,
$A_{\rm Airy}=3.35(\lambda/D)^2$ (for pupil obscuration $\epsilon=0.25$).
The sky noise is thus equivalent to that in a circular aperture of
radius $1.03\lambda/D$.  In the limit $P\gg1$, $A_{SN}$ is simply the
pixel area $A_{\rm pix}=P^2(\lambda/D)^2$.  It should be noted,
however, that the common heuristic approximation $A_{SN}\approx A_{\rm
Airy} + A_{\rm pix}$ will {\it underestimate} $A_{SN}$, and hence
the required exposure time, by up to 40\% between these limits.
In particular, note that Nyquist-sampling pixels ($P=0.5$)
degrade the Airy $A_{SN}$ level by 13\%, while pixels at the Airy FWHM of
$P=1.22$ degrade the speed by about a factor 1.5, assuming interlacing
to the Nyquist level.

The degradation of point-source $S/N$ is more severe if the detector
has significant diffusion of charge before collection into pixels.
The dashed line in Figure~\ref{ASNfig} shows $A_{SN}$ when there is
Gaussian charge diffusion with $\sigma$ of one-half pixel.  In this
case even the Nyquist-sized pixels increase $A_{SN}$ by 30\%, and the
$A_{SN}$ for $P=1$ is $\approx10$ pixels, three times worse than
the pure Airy value.

On the other hand, the overall speed of a photometry project can be an
increasing function of pixel size if the number of pixels is
constrained.  If the science goals require surveying a fixed, large
number of square degrees to a given depth, then the total time to
complete the project scales as
\begin{equation}
T \propto { A_{SN} \over N_{\rm pix} P^2 }.
\end{equation}
So for fixed pixel count $N_{\rm pix}$, the figure of merit is
$P^2/A_{SN}$, which we see from Figure~\ref{ASNfig} is always
increasing with $P$, to an asymptotic value of unity.  A grossly
undersampled camera at $P=10$ conducts a point-source survey 10 times
faster than a Nyquist-sampled camera!  In the presence of charge
diffusion, the undersampled camera is still $4\times$ faster.

For sources brighter than the sky background, the $S/N$ is independent
of pixel size, hence the survey speed grows with the FOV, or $\propto
P^2$ if $N_{\rm pix}$ is fixed---an even stronger advantage than the
background-limited case.

The gains of larger pixels are realized only as long as the FOV
increases linearly with pixel size.  In reality, aberrations and
engineering difficulties will place lower bounds on the focal length
and upper bounds on the FOV.  But these results suggest a very strong
motivation toward coarse pixels in space-based survey projects.
Coarse pixels can have other well-known practical advantages with
respect to dark current, read noise, and telemetry rates.

The above calculations assume PSF-fitting on isolated point sources.
This is wholly appropriate for time-domain projects (supernova
hunting, microlensing, moving objects) in which a high-S/N template
image can remove all but the (rare) variable objects from an image.
For other projects, however (crowded-field stellar photometry), the
large pixels will impose a severe crowding penalty, and $P\lesssim1$
will be strongly preferred.  A project for which morphological
information is essential will of course suffer with coarse pixels as
the high-frequency information is strongly attenuated when $P\gg1$.
Such projects really require optimization of image reconstruction
as discussed in L99a and \citet{HF00}.

Even time-domain projects will suffer significantly when the pixels
become large enough that the shot noise from neighboring (static)
objects begins to outstrip the ecliptic sky background in a typical
pixel.  This occurs when the pixel size is comparable to the typical
spacing between objects that have surface brightness above that of the
zodiacal light background.  At high galactic latitudes, stars are
rare, and galaxies with surface brightness above
23--24~mag~arcsec$^{-2}$ will be many arcseconds apart from
each other.  For nearby supernovae, the host galaxy's central surface
brightness may exceed the zodiacal light, so we favor small pixels which do
not blend the nucleus/center with the supernova.
Microlensing and stellar-variability surveys will typically point
toward nearby galaxies with many bright individual stars, so very
large pixels may increase the effective background level.

In summary, for survey-oriented projects there are very strong
efficiency gains from $P\gtrsim1$ if pixel count and/or telemetry
bandwidth are limited.  

\subsubsection{Point-Source Astrometry}
\label{astrometry}
\begin{figure}[t]
\plottwo{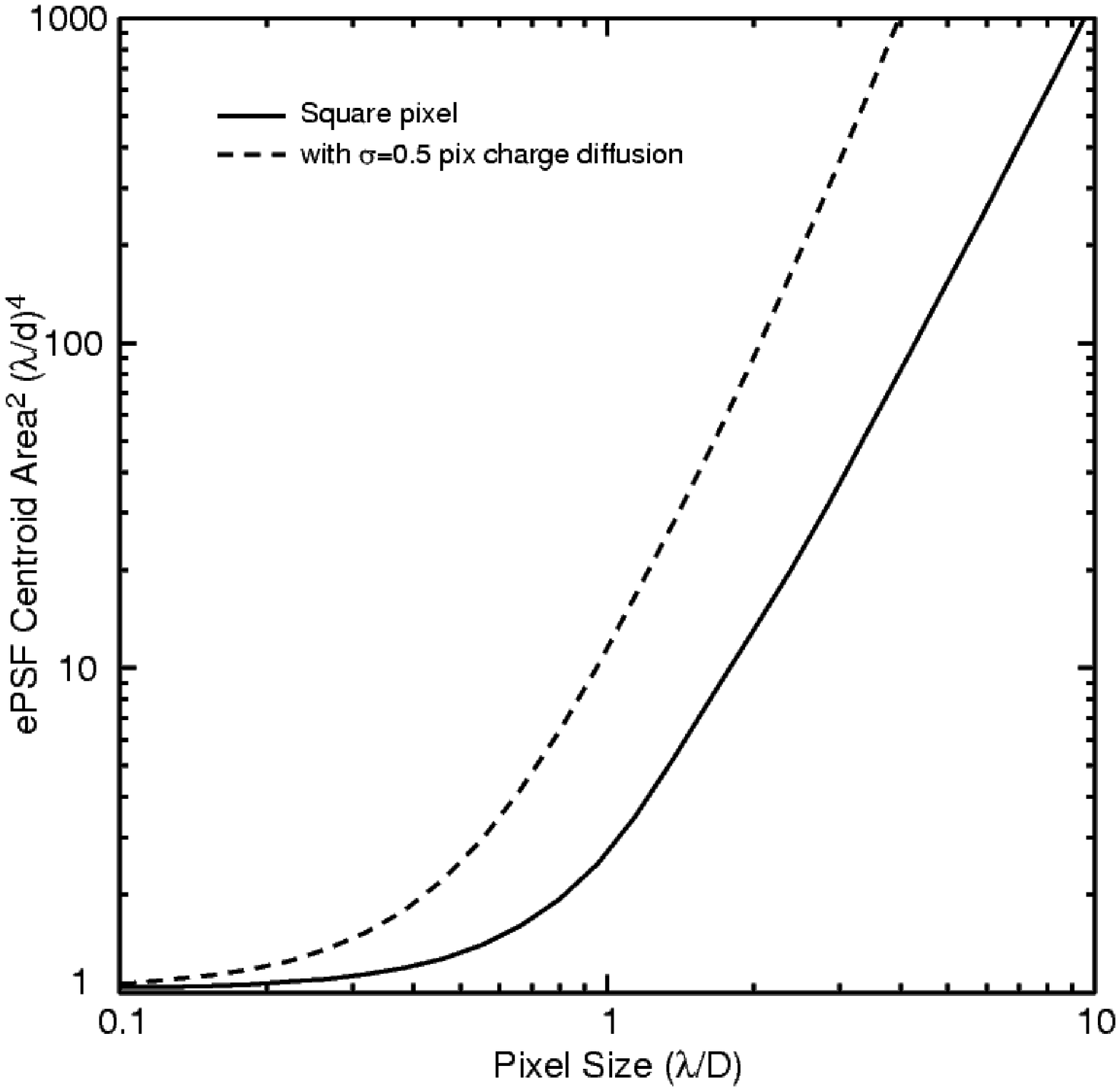}{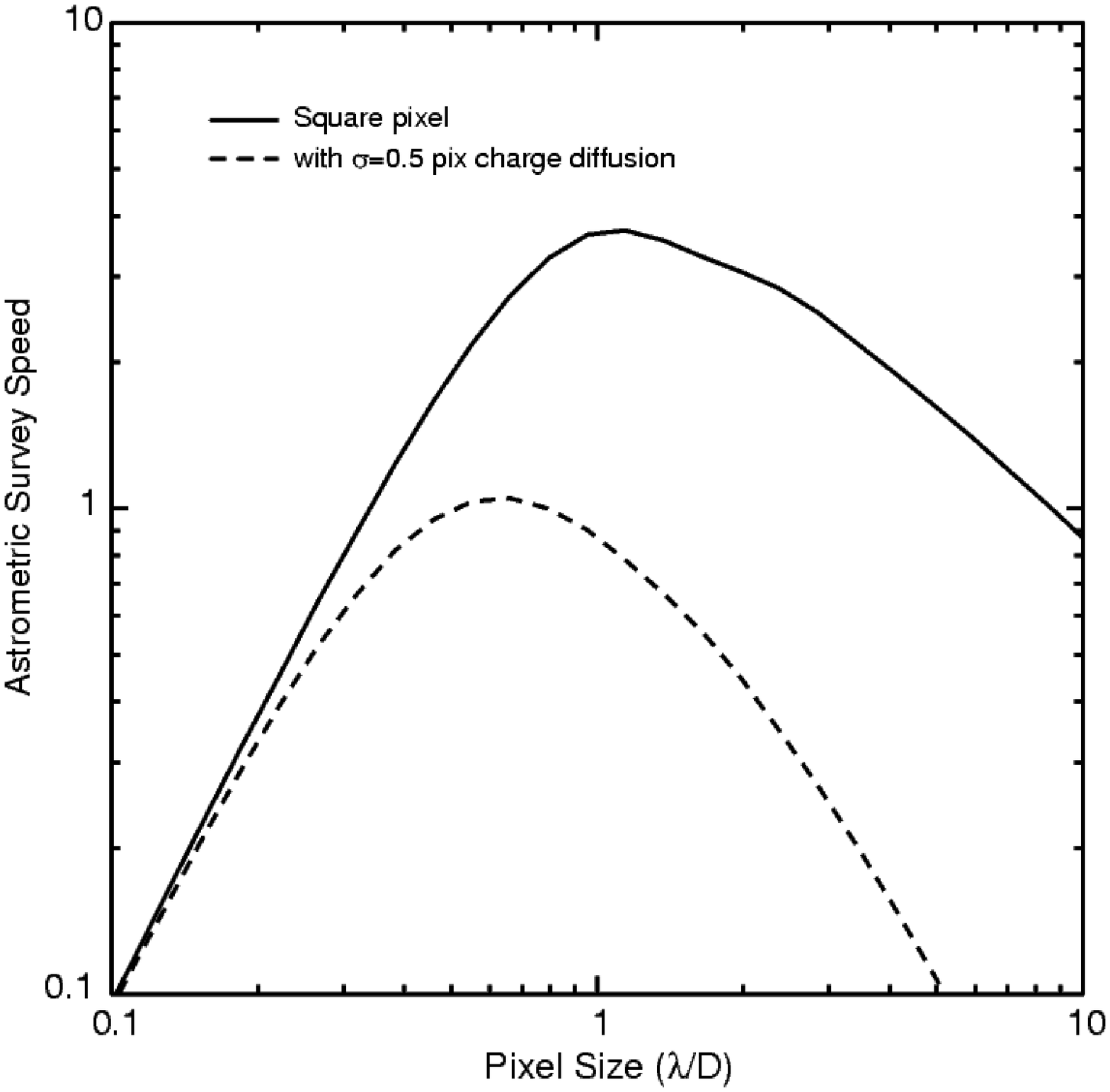}
\caption[]{
\small
On the left plot, the effective area $A^2_{\rm cent}$ of the ePSF for
background-limited point-source {\em astrometry}
is plotted vs pixel size.  The solid line is for a perfect
square pixel atop a diffraction-limited PSF from a telescope with 25\%
pupil obscuration.  The dashed line shows the effect of charge
diffusion with $\sigma=0.5$~pixels.  The time to obtain
fixed astrometric accuracy on a single source scales as $A^2_{\rm
cent}$. The penalties for large pixels and charge diffusion are more
severe than for photometry (Figure~\ref{ASNfig}).
The right side plots the
speed for an astrometric survey (sky coverage divided by time to a
obtain given centroid accuracy) vs pixel
size, given a fixed pixel count.  There is a clear optimum pixel scale
at 1--2 $\lambda/D$.
}
\label{astrompix}
\end{figure}

Astrometric measurements place a higher premium on compact ePSF than
do flux measurements since the centroid is a higher moment (first) of
the stellar image than is the flux (zeroth).  In
Figure~\ref{astrompix} I plot the relative survey speed for a
background-limited, 
Nyquist-sampled astrometric measurement as a function of pixel size,
again assuming a fixed $N_{\rm pix}$.  In this case there is a very
well-defined optimum size of $1\lesssim P \lesssim 2$.  Unlike the
flux-measurement case, there are no gains to larger pixels;
bright stars as well as background-limited stars will prefer
intermediate pixel scales.  The only reason to use Nyquist-sampled
pixels ($P\le0.5$) on an orbiting astrometric satellite
would be if there is no FOV gain from a coarser
scale.

\subsubsection{Galaxy Ellipticities}
\begin{figure}[t]
\plottwo{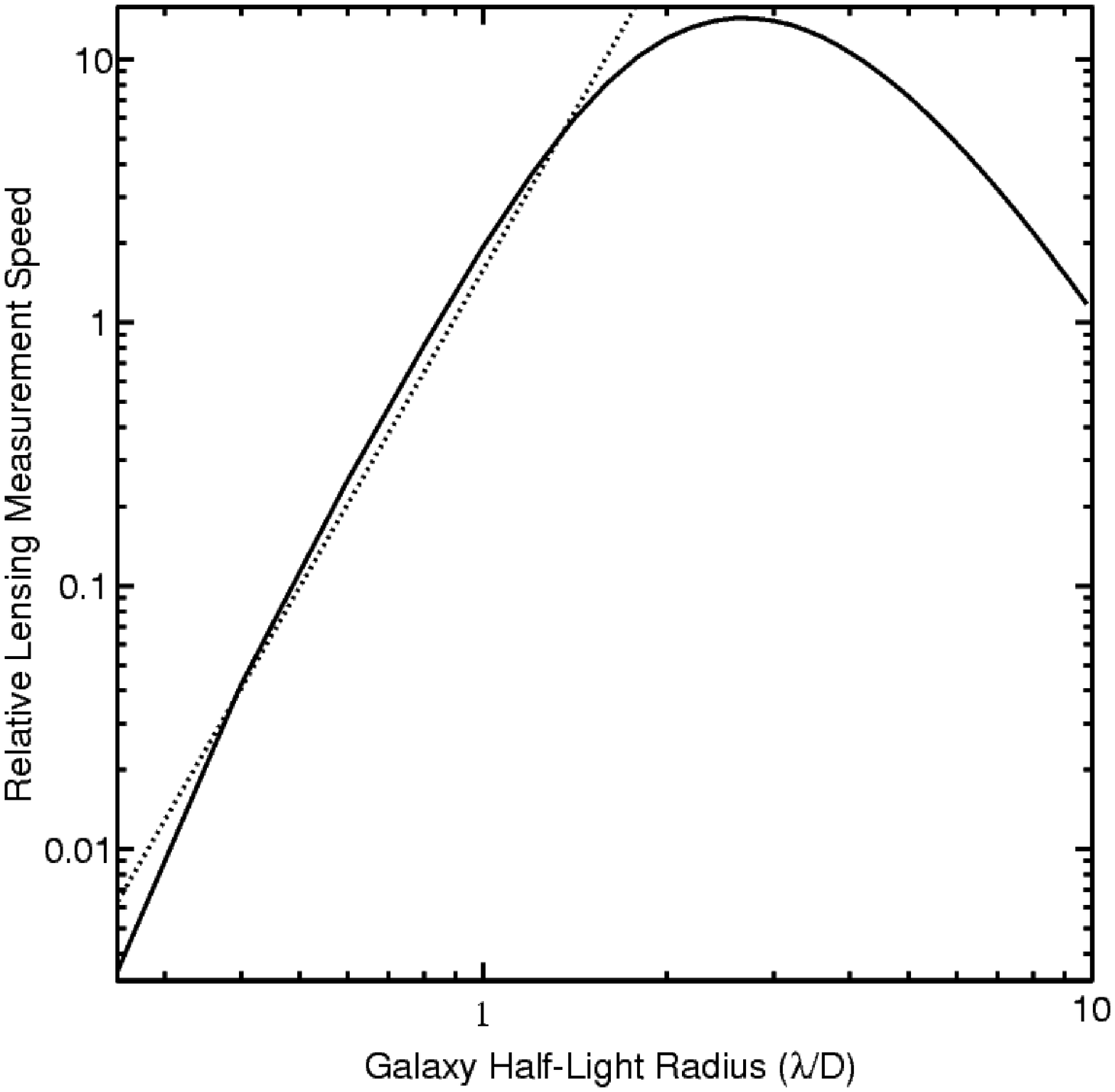}{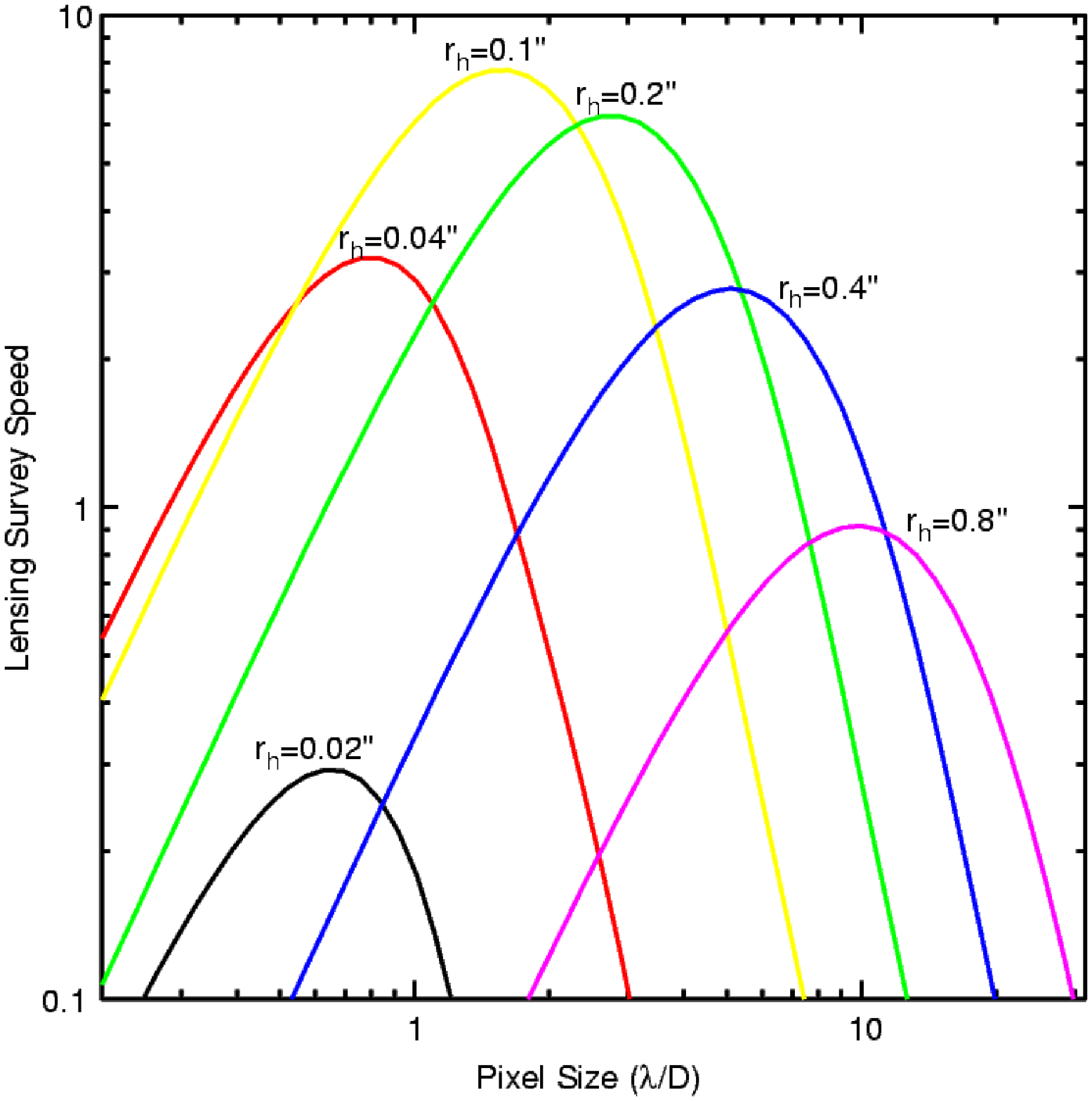}
\caption[]{
\small
On the left-hand side, 
the speed for a weak-lensing survey (inverse of time to reach
ellipticity accuracy of 0.2) is plotted versus the half-light radius $r_h$
of an assumed exponential-disk galaxy of fixed magnitude.
For this plot it is assumed that the pixels are of size $0.5\lambda/D$.
Large galaxies are slowly measured due to sky noise.  Poorly resolved
galaxies are strongly penalized:  the dotted line shows speed $\propto
r_h^{-4}$. 
On the right-hand side, lensing-survey speed
is plotted versus choice of pixel
size, given a fixed pixel count and galaxies of a given 
$r_h$.  All galaxies are given a common magnitude and are
assumed to be background-limited.  For this plot it is assumed that
$\lambda/D=0\farcs1$.  Galaxies with $r_h<0.4\lambda/D$ are poorly
measured at any pixel size; for measurable galaxies, the optimum pixel
sizes are in the range 0.8--3$\lambda/D$.  For galaxies that are many
times larger than $\lambda/D$, grossly oversized pixels are favored if
they come with increased FOV.
}
\label{lensfig}
\end{figure}

The optimization for weak-lensing observations depends upon the size
of the target galaxy.  Figure~\ref{lensfig} shows the time required to
reach $\sigma_e<0.2$ for a background-limited galaxy as a function of
the galaxy size $r_h$ at fixed galaxy magnitude.  When the galaxy is large,
the required time scales as $r_h^2$ since the galaxy is well-resolved,
but is spread over more background.  When the galaxy size $r_h$ is
below the ePSF size, the required time grows as a strong function
$\sim r_h^{-4}$ because the ellipticity information is suppressed by
the ePSF.  There is, hence, a significant penalty to making the pixel
scale too coarse, even if the FOV can be increased along with pixel
size.

This is quantified further in Figure~\ref{lensfig}, in which I plot
the relative speed vs pixel size $P$ for lensing measurements.  If $t$
is the time required to reach $\sigma_e<0.2$, then the survey speed is
$\propto N_{\rm pix} P^2 / t$.  The Figure plots relative lensing survey
speeds as a function of pixel size under the assumption of fixed
$N_{\rm pix}$.  In this case $\lambda/D=0\farcs1$, and the curves show
the speed for exponential-disk galaxies with $0\farcs02 \le r_h \le
0\farcs8$.  The vast majority of observable galaxies fall within this
range \citep{GS00,R98}.  The smallest and largest galaxies are
very poorly observed with this $\lambda/D$ regardless of pixel size.
For the intermediate sizes, we see that the optimal pixel scales are
$1 \lesssim P \lesssim 4$.  If $N_{\rm pix}$ is fixed, therefore, it is
once again advisable to make the pixels $\gtrsim \lambda/D$ in
size---{\em not} at the Nyquist size $0.5\lambda/D$.

\subsection{Comparison of Ground-Based and Orbiting Observatories}
Here I use the methods and software described above to compare the
survey capabilities of two proposed observatories:  {\em SNAP}
would represent the state of the art in 
orbiting imaging observatories late in the decade, with 
a 1~deg$^{2}$ CCD FOV behind a 2-meter telescope.  The {\em Large
Synoptic Survey Telescope}\footnote{{\tt
http://www.lssto.org}} ({\em LSST}) would likewise represent the
state of the art in large ground-based survey telescopes, with $\approx
7$~deg$^{2}$ FOV behind an 8.4-meter primary mirror.  In terms of
imaging throughput, each instrument would be $\approx2$ orders of
magnitude faster than present-day counterparts.  The space and ground
observatories, however, have very distinct strengths, and would likely
be focused on very different science goals.

The assumed characteristics of the two observatories are detailed in
Table~\ref{obschars}.  The important differences to note are:
\begin{itemize}
\item {\em SNAP} pixels are 0\farcs1, corresponding to $2\lambda/D$ for the
1~$\mu$m diffraction-limited PSF, whereas the {\em LSST} pixels are
0\farcs25 to sample the presumed 0\farcs5 FWHM ground-based seeing.
\item Dithering is assumed to be ineffective from {\em LSST} due to the
time-variable PSF.  An optimal interlacing factor is chosen for each
proposed {\em SNAP} observation.
\item {\em SNAP}, in high-earth orbit, is assumed to be on target nearly
full time.  For {\em LSST} I presume that on average only 30\% of the time
is useful after losses due to daylight, moonlight, and clouds.
\item {\em LSST} has a somewhat larger pixel count, and is assumed to have
faster readout (5s vs 20s).
\item The {\em LSST} secondary obscuration is quite large (55\% of the
primary aperture) compared to {\em SNAP} and typical 2-mirror telescopes.
\item The {\em SNAP} background is taken to be the zodiacal brightness at
the North ecliptic pole, while the {\em LSST} sky is the new-moon Cerro
Tololo zenith sky brightness.  {\em LSST} zenith atmospheric
extinction is taken to be that at Cerro Tololo as well.
\end{itemize}

\begin{deluxetable}{lcc}
\tablewidth{0pt}
\tablecaption{Assumed Observatory Characteristics}
\tablehead{
\colhead{Quantity} &
\colhead{{\em SNAP} Value} &
\colhead{{\em LSST} Value}
}
\startdata
Telescope Aperture & 2.0 m & 8.4 m \\
Focal Length	& 21.6 m & 8.2 m \\
Fractional Diameter of Pupil Obscuration & 20\% & 55\% \\
Gaussian $\sigma$ for Aberrations/Seeing & 0\farcs05 & 0\farcs5 \\
Optical Transmission & 83\% (Vis \& NIR), 50\% (UV) & 70\% of zenith
atmosphere\tablenotemark{1} \\
CCD Pixel Size & 10.5 \micron & 10.5 \micron \\
CCD Read Noise & 4 e & 4 e \\
CCD Quantum Efficiency & LBL CCD\tablenotemark{2} & LBL
CCD\tablenotemark{2} \\
CCD Charge Diffusion Sigma & 3.5 \micron & 3.5 \micron \\
CCD Dark Current & 0.0013 e~s$^{-1}$~pix$^{-1}$
& 0.0013 e~s$^{-1}$~pix$^{-1}$ \\
CCD Readout Time & 20 s & 5 s \\
CCD Cosmic-Ray Rate & 0.00013 s$^{-1}$~pix$^{-1}$ & 0 \\
CCD FOV & 1.0 deg$^2$ & 7.0 deg$^2$ \\
NIR Pixel Size & 18 \micron & 18 \micron \\
NIR Read Noise & 4 e & 4 e \\
NIR Quantum Efficiency & HgCdTe\tablenotemark{3}
& HgCdTe\tablenotemark{3} \\
NIR Charge Diffusion Sigma & 5 \micron & 5 \micron \\
NIR Dark Current & 0.02 e~s$^{-1}$~pix$^{-1}$
& 0.02 e~s$^{-1}$~pix$^{-1}$ \\
NIR Readout Time & 1 s & 1 s \\
NIR $N_{\rm pix}$ & $6.4\times10^7$ & $6.4\times10^7$ \\
Sky Brightness & Ecliptic Pole Zodiacal\tablenotemark{4}
 & CTIO Zenith Dark Sky\tablenotemark{5} \\
Duty Cycle & 100\% & 30\% \\
\enddata
\tablenotetext{1}{Atmospheric extinction from \citet{Ha92,Ha94}}
\tablenotetext{2}{Expected QE for High-Resistivity CCD used, \citep{Gr00}}
\tablenotetext{3}{Measured QE for existing HgCdTe HAWAII arrays}
\tablenotetext{4}{Zodiacal brightness from \citet{Le97}}
\tablenotetext{5}{CTIO zenith sky brightness from \citet{Ma00}}
\label{obschars}
\end{deluxetable}

For wavelengths beyond 1~$\mu$m, I posit either {\em LSST} or {\em SNAP}
to be equipped with a mosaic of 16 2k$\times$2k$\times15\mu$m HgCdTe
array detectors, with 4e read noise and 0.02 e/s dark current.  The
HgCdTe pixels are assumed to have a 10\% dead zone on each edge.
I presume for now that the NIR arrays would have
the same focal ratio and cosmic-ray rates as the posited CCD arrays.

\subsubsection{Point-Source Photometric Survey}
\begin{figure}
\plotone{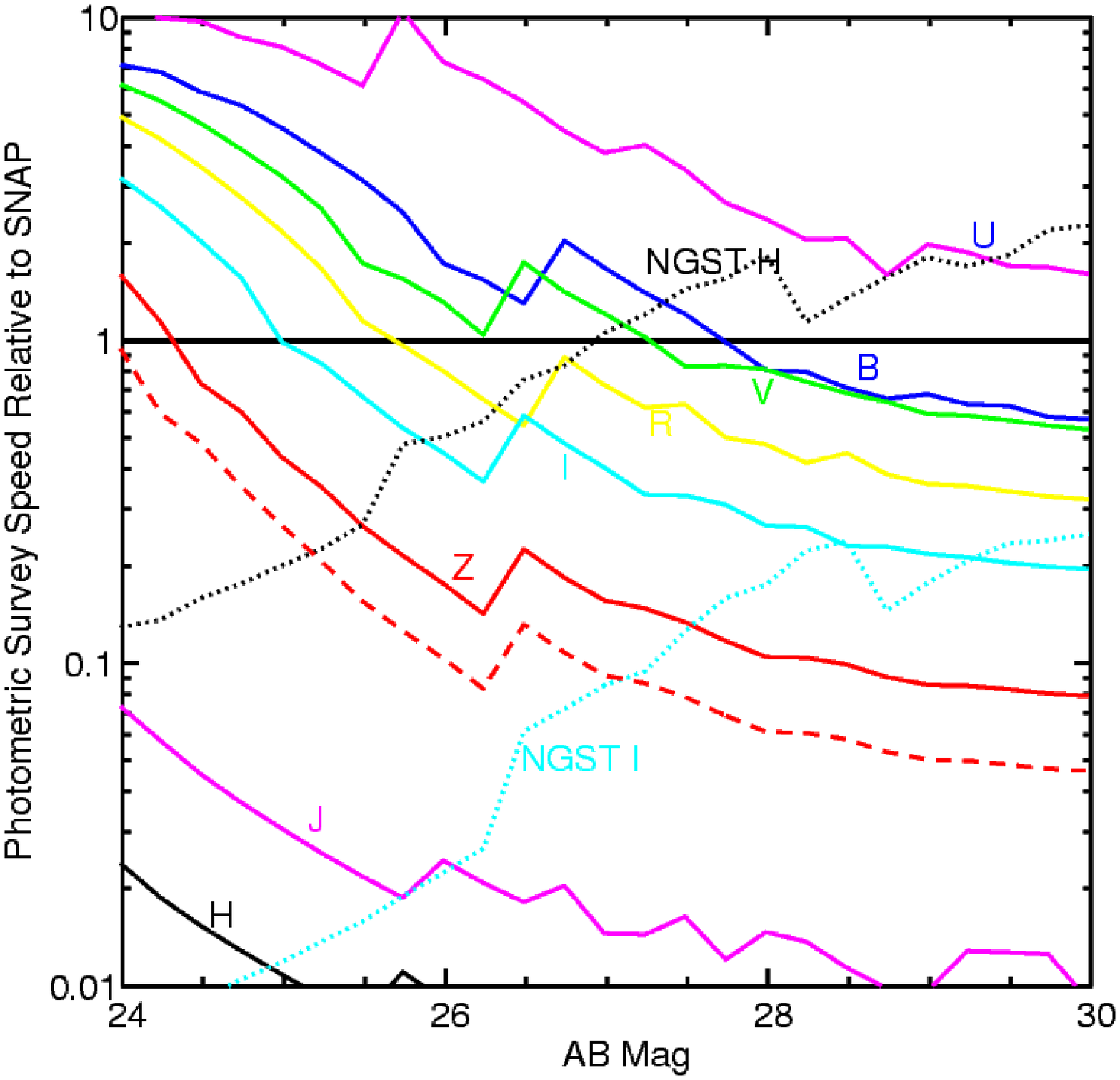}
\caption[]{
\small
The survey speed for photometric observations (sky coverage divided by
time to fixed $S/N$) is given for {\em LSST} relative to {\em SNAP}.  Different
filter bands are labeled.  The dashed red curve shows the result of
having 0\farcs7 FWHM seeing instead of the 0\farcs5 (solid red) in $Z$
band.  Poorer seeing would shift all other wavebands similarly.  Two
curves are shown for the speed of an NGST NIR/visible 8k imager relative
to {\em SNAP}.  I demand in each case that 95\% of point sources be
measured to $S/N\ge7$; the jumps in the curves occur where cosmic-ray
hits become likely in $>5\%$ of pixels.
}
\label{ptsrc}
\end{figure}

Figure~\ref{ptsrc} compares the speed for a photometric point-source
survey on {\em LSST} relative to {\em SNAP}.  The figure of merit being compared
is the number of square degrees of sky per 24-hour period
which can be surveyed for given source magnitude.  I demand that at
least 95\% of the sources at the chosen magnitude be measured with
$S/N\ge7$ (recall that cosmic rays and varying pixel phases make the
$S/N$ a random variable).

We can reach the following conclusions:
\begin{itemize}
\item When the sources have $AB<27$ and are observable in $B$ and $V$
bands ({\it e.g.} a low-redshift 
supernova search), the ground-based search is more
efficient by a factor of 2--5.  
\item For $R$-band observations, or for very faint $B$-band sources, there
    is no clear advantage.
\item When the sources move to $I$ band and $AB>27$, the faint background
    and resolution of the space platform start to win.  $I$-band surveys
    are 3--5 times faster from orbit.

\item In $Z$-band the space advantage is 7--10 times.

\item In the NIR the space advantage is of course huge due to background
    issues. In $J$, the 2m {\em SNAP} is 30--100 times faster than the
8.4m {\em LSST}, given a comparable investment in IR array detectors.
The $H$ band advantage is even larger.
\item Also shown
on the plot are the relative figures of merit for the proposed
8k$\times8$k NIR/visible imager aboard the 8-meter {\it Next
Generation Space Telescope} (NGST).  In the NIR, the larger aperture
makes NGST $\approx2\times$ faster than {\em SNAP}, but only for the
very faintest sources observable by {\em SNAP}.
In $I$-band, the much larger FOV of {\em SNAP} makes the survey 
at least $3\times$ faster.
\end{itemize}

This of course is just a noise analysis; there are systematic-error
and cost issues as well, the former favoring {\em SNAP} and the latter
{\em LSST}.  In particular, note that the above analysis has assumed
isolated point sources---which is appropriate to a time-domain search
with perfect difference imaging.  For crowded-field photometry,
however (color-magnitude diagrams for distant systems, Cepheid
measurement, etc.), the space telescopes gain a large factor from
better resolution.

Note also that both observatories' designs could be tweaked to improve
performance on this measure, but the ultimate restrictions on
FOV, telemetry rate, etc. require a full engineering analysis.

\subsubsection{Photometric Redshifts}
For lensing applications, pre-determination of supernova host-galaxy
redshifts, and a 
slew of galaxy-evolution studies, photometric determination of
galaxy redshifts will be of huge benefit.  We thus need to know the
speed at which we can measure colors of resolved objects to a nominal
accuracy.  I take here a target $S/N\ge20$ for photo-z applications.

\begin{figure}
\plotone{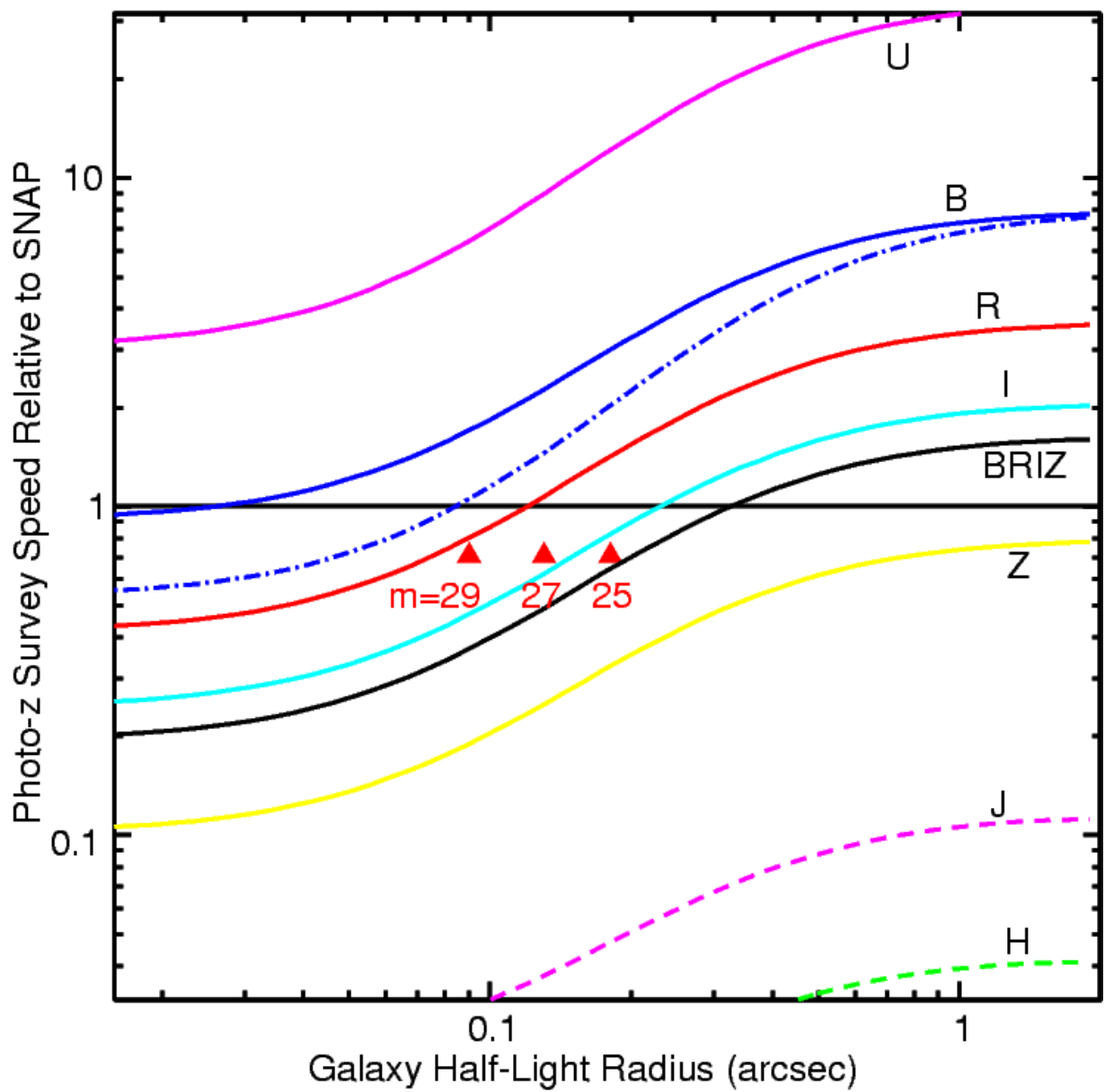}
\caption[]{
\small
The survey speed for galaxy photometric redshift surveys (sky coverage
divided by time to fixed flux error in Gaussian apertures) is given
for {\em LSST} relative to {\em SNAP}, for various wavelengths.  Typical
faint-galaxy sizes are marked.  The ground observations are favorable
only for larger galaxies at bluer wavelengths, as expected. 
}
\label{gspeed}
\end{figure}

Figure~\ref{gspeed} plots show the relative speeds of the nominal {\em SNAP} and {\em LSST}
configurations for photometry of galaxies.  Here it has been assumed
that fluxes of galaxies are being measured through Gaussian apertures,
and the aperture size has been selected to optimize the S/N.  This is
close to an optimal procedure for the exponential-disk galaxies
considered here.  Poisson noise from sky, source, and dark counts is
included.  Read noise is negligible.  The plots are for $AB=27$ mag
galaxies of various sizes, but the relative speeds will apply to any
source that is fainter than the background.
Again an optimistic duty cycle of 30\% has been assumed for the
ground-based survey.

The relative speed is plotted as a function of galaxy half-light
radius.  The typical sizes for faint galaxies found in the HDF-South
STIS images are marked with triangles (from \citet{GS00}).  Galaxy
photometry is seen to be faster from the ground in the blue, but a
good deal faster from space for $Z$, $J$, and $H$-band observations,
with a near tie in $R$ and $I$ bands.  
The black line shows the relative speed when all four visible
bands $B$, $R$, $I$, \& $Z$ are to be done sequentially.  

For nearby galaxies with sizes $\gtrsim0\farcs5$, any CCD-based
photometric redshift survey is better done from the ground.  But the
orbiting observatory wins heavily in the NIR bands, or for the smaller
galaxies more typical at $m_{AB}\gtrsim27$.

No flat-fielding errors or crowding have been included here.  The
latter will be important for ground-based images at 29--30~mag.

\subsubsection{Ellipticity Measurements}
\begin{figure}

\plotone{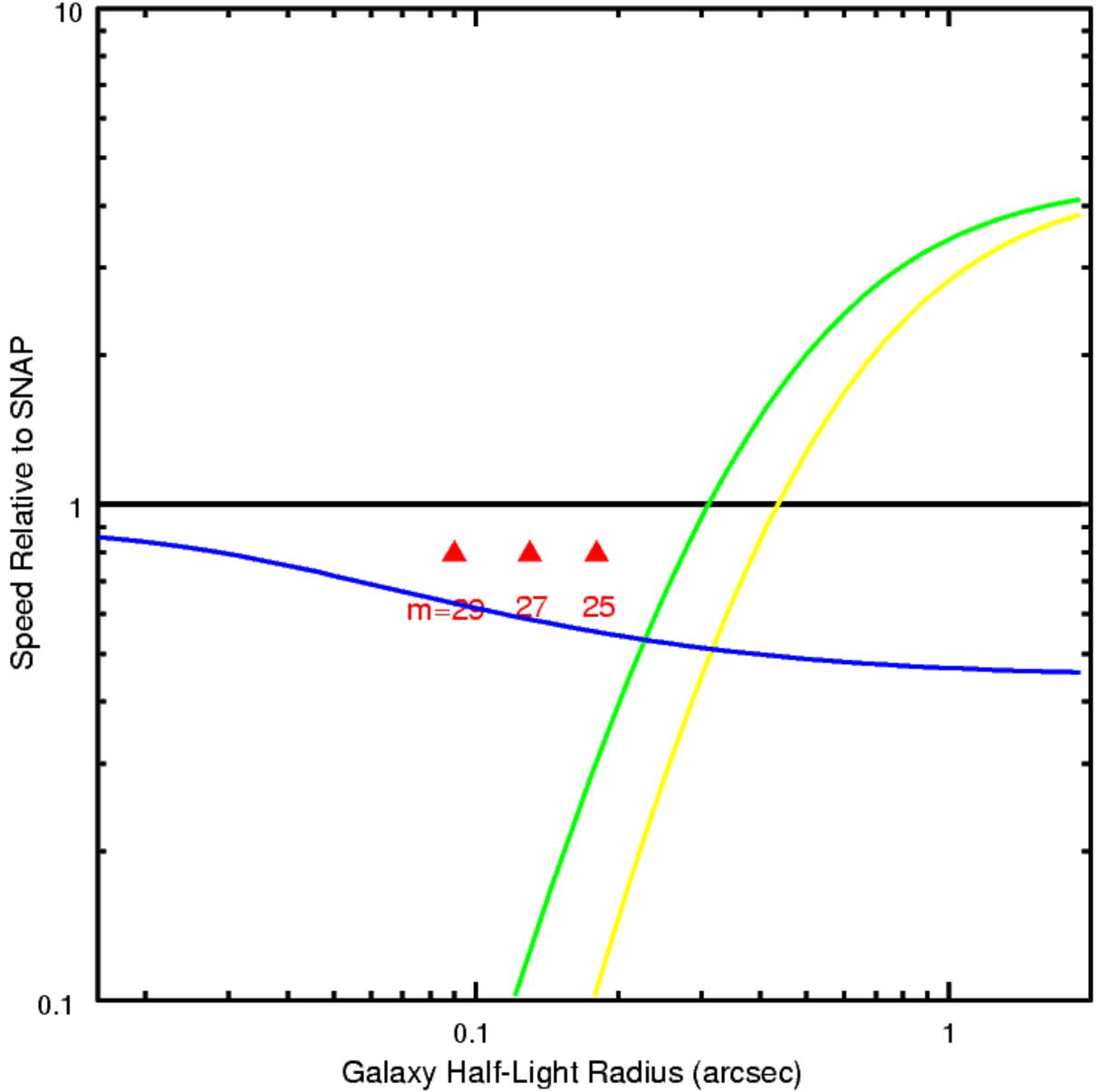}
\caption[]{
\small
The survey speed for weak lensing observations (sky coverage divided by
time to fixed ellipticity error at $m_{AB}=27$) is given for {\em LSST}
relative to {\em SNAP}.  
{\em SNAP} observations are assumed to be in $I$ band, {\em LSST} in $R$ band, and
all observations are assumed to be background-limited with perfect
systematic-error elimination.  The green (yellow) curves assume
Gaussian seeing with 0\farcs5 (0\farcs7) FWHM.  The typical sizes of
faint galaxies in the HDF-S are marked.  For such galaxies, {\em SNAP} has a
modest to large speed advantage, whereas large galaxies are best done
from the ground.  The blue curve compares a {\em SNAP} $B$-band
observation to the nominal $I$-band, assuming a typical galaxy color.
}
\label{espeed}
\end{figure}

Figure~\ref{espeed} shows the relative speeds for lensing observations.
In each case, the figure of merit is the time it takes for the background
noise to be reduced to the point where the galaxy
ellipticity is measured to an accuracy of 0.2 or better.  The galaxy
is assumed here to have $m_{AB}=27$, and
be a circular exponential disk; the relative speeds will remain the
same for any background-limited case.
Overhead and cosmic-ray hits are ignored, as is appropriate for deep images.

A lensing survey can be conducted in the filter of choice (apart from
a desire for photometric redshifts, described above).  One would
likely choose something like $I$-band for a space observation and $R$
band from the ground to obtain the shape information most rapidly.
From the figure it is clear that {\em LSST} surveys large galaxies about
4--5 times faster than {\em SNAP}, due to the larger aperture and FOV,
if I assume fixed $m_{AB}$ vs wavelength.
In actuality most galaxies are redder
than this, so the relative speed of {\em LSST} will be decreased by about
$2\times$ (if $\langle R_{AB}-I_{AB} \rangle\approx0.4$~mag).

For galaxies smaller than 0\farcs3 half-light radius, {\em SNAP} is faster
because the ground-based seeing 
dilutes the signal and squelches the ellipticity signal.  Indeed the
required exposure times rise very rapidly ($\propto r^{-4}$) when
galaxies are poorly resolved, so in fact an orbiting observatory is
essentially required to extract lensing information the bulk of galaxies at
$m_{AB}\gtrsim25$.  High-order adaptive optics are never likely to
cover sufficient FOV to make practical weak-lensing observations, but
the wide-field tip-tilt-correction scheme of {\em WFHRI} \citep{KTL} may allow
ground-based weak-lensing observations of smaller galaxies.  I hope to
analyse the relative merits of the {\em WFHRI} configuration in the near
future; my expectation (and that of \citet{KTL}) is that {\em WFHRI} will be
faster for $BVR$ observations but the low background of {\em SNAP} will win
out in $Z$ or $I$.  
I have completely neglected {\it systematic} errors in correcting galaxy
shapes for PSF anisotropies.  Eliminating these errors will certainly
be much more difficult in ground-based images, both {\em LSST} and {\em WFHRI}
type, because the PSF will
have strong variation in both space and time.  The density of point
sources must be high enough to be able to track these variations to
the desired accuracy.

\section{Summary}
The tools presented herein, while not particularly original or clever,
permit one to optimize hardware and observing-protocol designs while
accounting for effects that are typically ignored in
aperture-photometry exposure-time estimates.  Proper consideration of
pixelization effects and cosmic-ray hits can easily change the
expected $S/N$ levels by a factor of 2, for example.  With this
machinery in hand, I have addressed a few issues of general interest,
such as quantifying the effects of ``oversized'' pixels on
photometric, astrometric, and lensing measurements, and showing that
interlacing exposures in a $3\times3$ pattern will extract essentially
all the useful information.  The phase space of observations
for which orbiting imagers are advantageous has been delineated as
well.  More importantly, the tools presented here are very general and
flexible, and can be applied to a great variety of future design
optimizations. 

\acknowledgements
This work was supported by grant AST-9624592 from the National Science
Foundation, and by DOE grant DE-FG-02-95-ER-40899.  Thanks to Alex Kim
and the members of the {\em SNAP} collaboration for having collected
most of the relevant data.

\end{document}